\documentclass[twocolumn]{aastex631}

\usepackage{amsmath}
\usepackage{graphicx}
\usepackage{subfigure}
\usepackage{paralist}
\usepackage{amssymb}
\usepackage{bm}
\usepackage{bbding}
\usepackage{multirow}
\usepackage{footmisc}
\shorttitle{Resonant network}
\shortauthors{Zhang \& Lei}

\begin{document}

\title{Resonant Networks of Spin--Orbit Coupling in Ellipsoid-Ellipsoid Binary Asteroid Systems}

\correspondingauthor{Hanlun Lei}\email{leihl@nju.edu.cn}

\author{Yuanzhe Zhang}
\affiliation{School of Astronomy and Space Science, Nanjing University, Nanjing 210023, China}
\author{Hanlun Lei}
\affiliation{School of Astronomy and Space Science, Nanjing University, Nanjing 210023, China}
\affiliation{Key Laboratory of Modern Astronomy and Astrophysics in Ministry of Education, Nanjing University, Nanjing 210023, China}



\begin{abstract}
The dynamical evolution of binary asteroid systems is deeply influenced by spin--orbit resonances. However, their domains of influence and mutual interactions remain elusive, in particular in the space where multiple resonant modes coexist. In such regimes, the standard single-resonance approach is intrinsically limited and fails to capture the true coupled dynamics. To overcome this, we develop a global Hamiltonian framework based on elliptic expansions of the spin--orbit coupling model, enabling the numerical construction of comprehensive resonant networks. Concentrating on a representative synchronous region that encompasses synchronous spin--orbit, spin--spin, spin--orbit--spin, and doubly synchronous resonances, we study the dynamical boundaries of different resonant modes in a systematical manner. Crucially, we identify a secondary resonance structure arising from the strong nonlinear coupling between the synchronous resonances of the primary and secondary asteroids. Ultimately, this study provides a reliable parameter-space atlas, which is helpful for predicting the long-term evolutionary pathways of binary asteroid systems.
\end{abstract}


\keywords{Perturbation methods (1215) --  Astronomical simulations (1857) -- Asteroid dynamics (2210) -- Orbital resonances (1181)} 


\section{Introduction} 

In celestial mechanics, spin--orbit resonance is a fundamental dynamical phenomenon where a celestial body’s rotational and orbital periods are locked in a commensurable ratio \citep{goldreich1966spin}. A classic example is the Moon, which 
exhibits synchronous rotation with Earth \citep{peale1969generalized}, while Mercury occupies an asynchronous 3:2 spin--orbit resonance, completing three rotations for every two orbits \citep{peale1965rotation, lemaitre2006spin}. With the discovery of binary asteroid systems \citep{margot2002binary, margot2015asteroid}, the study of spin--orbit resonances can be naturally extended to these full two-body environments \citep{scheeres2006dynamical}. Furthermore, recent mission-oriented studies—evaluating gravitational field models for landing and designing goal-directed guidance strategies—depend heavily on the accurate characterization of these resonant dynamics to ensure safe and efficient proximity operations \citep{wen2024comparative, rizza2025goal}.

When studying these resonances, classical models often assume a fixed Keplerian mutual orbit \citep{goldreich1966spin, celletti1990analysis2}. This approximation is justified when the orbital angular momentum is significantly greater than the rotational angular momentum, as the effects of angular momentum exchange on orbital motion can be safely neglected \citep{celletti1990analysis1}. Within this framework, the stability of spin--orbit resonances and the influence of orbital elements have been systematically studied, together with the development of dedicated analytical and numerical methodologies.  
  Specifically, \citet{wisdom1984chaotic} attributed Hyperion's chaotic tumbling to the resonant dynamics of its triaxial shape, and \citet{celletti2000hamiltonian} clarified the role of eccentricity in selecting and stabilizing these resonances. Moreover, \citet{jafari2015widespread} highlighted the widespread presence of chaotic rotation for the secondary body in a binary asteroid system, and \citet{seligman2021onset} adopted a two-dumbbell model  to analyze the onset of chaotic rotation arising from the overlap of spin--orbit and spin--spin resonances. Parallel to these developments, analytical theories in celestial mechanics have found wide application to the spin--orbit problem. In this respect, \citet{flynn2005second} extended Lie transform perturbation theory to the second order, 
characterizing its region of validity for the planar spin-orbit problem. 
Subsequently, \citet{celletti2008measures} introduced a method for computing the measures of basins of attraction  for resonances in a dissipative spin--orbit model, quantifying the fraction of initial conditions evolving to different spin–orbit attractors.  More recently, \citet{deoliveira2025multistability} introduced the Gibbs entropy as an effective metric to quantify the uncertainty of the system's final resonant state. Building on the classical Keplerian framework, the study of novel resonance types has been facilitated, as demonstrated by \citet{gkolias2016theory}, who provided a high-precision analytical prediction of the existence and stability of secondary resonances within the synchronous region. Furthermore, \citet{Celletti2022_SpinSpin} distinguished between the standard and balanced resonances, and investigated the boundary conditions leading to symmetric periodic orbits and uncovered a phenomenon of measure synchronization in spin--spin resonances. Under the theoretical framework established by \citet{wisdom2004spin}, \citet{callegari2024hamiltonian} explored the 1:1 secondary resonance in satellites of giant planets to detail the coupling between the free frequency of physical libration and the orbital mean motion.

For binary asteroid systems characterized by 
close separations, the orbital and spin angular momenta become comparable in magnitude. Consequently, the exchange of angular momentum actively drives the evolution of the mutual orbit. Capturing these complex interactions requires a fully coupled spin--orbit model, a framework that has significantly advanced our understanding of coupled resonant dynamics and unveiled novel resonance configurations. Leveraging this approach, \citet{naidu2015near} developed numerical simulations to analyze the coupled spin dynamics of near-Earth Asteroid (NEA) satellites. Subsequently, \citet{hou2017note} established a criterion to determine how the equilibrium points of spin--orbit, spin--spin, and spin--orbit--spin resonances shift in response to the mutual separation. Regarding internal resonance structures, \citet{wang2020secondary} identified an anticorrelation between maximum libration amplitude and orbital eccentricity within secondary resonances, and \citet{jafari2023surfing} introduced a dynamical proximity criterion to explore how system parameters reshape these structures and induce chaos. Recently, \citet{lei2024primary} applied an adiabatic approximation to resolve spin--orbit resonances near the critical semi-major axis—a regime where the traditional pendulum approximation breaks down—and \citet{lei2024secondary} employed advanced perturbative techniques to probe the rotational dynamics of secondary components in these binary systems.

Within the framework of spin--orbit coupled dynamics, although various resonances have been identified and their individual properties characterized, 
a critical question arises: how to predict the final resonance state of a binary asteroid when multiple resonances coexist within a parameter space. The traditional approach of analyzing resonances in isolation is effective outside these complex regions; within them, however, one must discern where the single-resonance approximation remains applicable and where a broader, multi-resonance framework is required. This necessity stems from the concurrent dominance of multiple resonances and the suppression of high-order ones by 
dominant low-order ones, which can 
alter the system’s evolutionary path and render single-resonance predictions unreliable. Therefore, a global and multi-resonance 
framework is essential to accurately delineate the dynamical structures of these interacting resonances.

Achieving such a predictive and globally self-consistent delineation remains a challenge. The importance of specifying resonance domains was highlighted by \citet{hou2017note}, who emphasized that spin--orbit--spin and spin--spin resonances should be defined outside the synchronous regions. However, a unified framework capable of 
incorporating these disparate resonance modes to systematically map their respective domains is still lacking.

In this work, we numerically construct the spin--orbit resonant networks for binary asteroid systems to 
establish the foundation for our analytical framework. Building on these networks, we focus on a representative region near the doubly synchronous resonances of both asteroids. Guided by the global dynamical structures, we systematically partition the resonance domains and analyze their individual dynamics. Our work 
yields a comprehensive overview, 
thereby enabling the prediction of evolutionary outcomes for binary asteroids when tidal dissipation is accounted for. Most notably, we identify a secondary resonance structure within the doubly synchronous region—a feature that 
eludes the traditional single-resonance model.

This paper is structured as follows. In Sect.~\ref{Sect2}, we briefly formulate the dynamical model for binary asteroid system based on elliptic expansions. The second-derivative-based index is introduced in Sect.~\ref{Sect3} to produce resonant networks of spin--orbit coupling. In Sect.~\ref{Sect4}, we partition the synchronous region into three-type domains, and elaborate on the dynamics in each domain. 
We apply the dynamical framework to a binary asteroid system (90) Antiope in Sect.~\ref{Sect5}. 
Finally, conclusions are summarized in Sect.~\ref{Sect6}.

\section{Dynamical model}
\label{Sect2}

\begin{figure}
    \centering
    \includegraphics[width=1\linewidth]{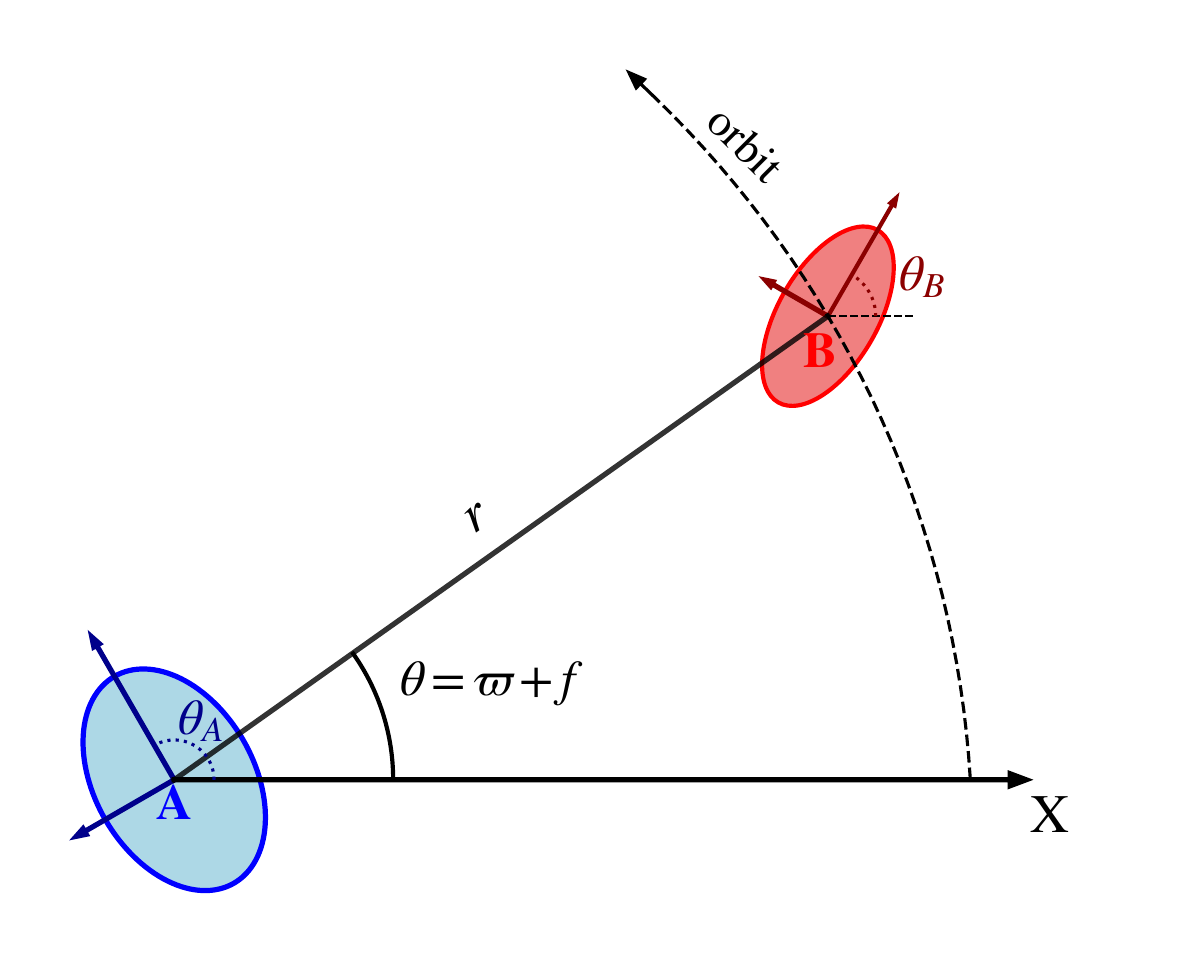}
    \caption{Schematic diagram of the relative geometry for an ellipsoid--ellipsoid binary asteroid system. The reference frame is centered on the primary asteroid ($A$). The secondary ($B$) is located at a distance $r$ from $A$, with the true longitude given by $\theta=\varpi+f$. The rotation angles of $A$ and $B$ are denoted as $\theta_A$ and $\theta_B$, respectively.}
    \label{Fig1}
\end{figure}

In this work, we consider a binary asteroid system, where the two asteroids move around each other under their mutual gravity field. To simplify this model, the following assumptions are made \citep{hou2017note,jafari2023surfing}:
\begin{itemize}
\item Both asteroids are modeled as triaxial ellipsoids. The primary and secondary are designated as $A$ and $B$, with masses $m_A$ and $m_B$, respectively. Their semiaxes satisfy 
\begin{equation}\label{Eq1}
    \begin{aligned}
        a_i\geq b_i \geq c_i>0, \; (i = A, B).
    \end{aligned}
\end{equation}
\item Both asteroids rotate about their respective axes of maximum inertia. The principal moments of inertia for each ellipsoid are given by
\begin{equation}\label{Eq2}
    \begin{aligned}
        \left\{ I^i_1, I^i_2, I^i_3 \right\} &= \frac{m_i}{5} \left\{
            \begin{aligned}
                &b_i^2 + c_i^2 \\
                &a_i^2 + c_i^2 \\
                &a_i^2 + b_i^2
            \end{aligned}
        \right\},\;
        (i = A, B).
    \end{aligned}
\end{equation}
\item The mutual orbit of the binary asteroid is confined to a plane, perpendicular to the asteroid's rotational axis. This means that the spin axis is aligned with the orbital angular momentum vector.
\end{itemize}
The relative geometry of the binary asteroid is shown in Fig.~\ref{Fig1}, where the coordinate system is centered at the barycenter of the primary asteroid $A$. The translation state of the secondary is represented by $(r,\theta)$, where $r$ is the radius and $\theta=f + \varpi$ is the true longitude. The rotation state is shown by $\theta_i$ and $\dot \theta_i$ ($i=A,B$). Under the ellipsoid--ellipsoid binary asteroid system, spin--orbit coupling implies that both the orbit and rotation are in coupled evolution subject to the conservation of mechanical energy and angular momentum of the system.

Under the aforementioned assumptions, we adopt the  the fourth order and degree (4OD) mutual potential \citep{hou2017mutual}. With the 4OD potential, we could formulate the system’s Hamiltonian in the following form:
\begin{equation}\label{Eq3}
\mathcal{H} = T + U,
\end{equation}
where $T$ is the kinetic energy, given by \citep{jafari2023surfing}
\begin{equation}\label{Eq4}
T = \frac{1}{2} m_{\rm AB} \bigl( \dot{r}^2 + r^2 \dot{\theta}^2 \bigr) 
      + \frac{1}{2} I_3^A \dot{\theta}_A^2 + \frac{1}{2} I_3^B \dot{\theta}_B^2
\end{equation}
with $m_{\rm AB}=\frac{m_A m_B}{m_A + m_B}$ as the reduced mass, and the potential function $U$ is truncated at the 4OD level, given by \citep{hou2017note}
\begin{equation}\label{Eq5}
U = -\mathcal{G} m_A m_B \Bigl( \frac{1}{r} + \frac{V_2}{r^3} + \frac{V_4}{r^5} \Bigr)
\end{equation}
where 
\begin{equation*}
V_2 = A_1 + A_2 \cos(2\theta - 2\theta_A) + A_3 \cos(2\theta - 2\theta_B),
\end{equation*}
and
\begin{equation*}
\begin{aligned}
V_4 =& B_1 + B_2 \cos(2\theta - 2\theta_A) + B_3 \cos(4\theta - 4\theta_A) \\
&+ B_4 \cos(2\theta - 2\theta_B)+ B_5 \cos(4\theta - 4\theta_B)\\
     &+ B_6 \cos(2\theta_A - 2\theta_B)+ B_7 \cos(4\theta - 2\theta_A - 2\theta_B).
\end{aligned}
\end{equation*}
The expressions of the coefficients $A_i$ ($i=1,2,3$) and $B_i$ ($i=1,2,...,7$) are given by \citep{hou2017note}
\begin{equation*}
\begin{aligned}
A_{1} &= -\frac{1}{2}(a_{A}^{2}C_{20}^{A}+a_{B}^{2}C_{20}^{B}), 
A_{2} = 3a_{A}^{2}C_{22}^{A}, 
A_{3} = 3a_{B}^{2}C_{22}^{B}, \\
B_{1} &= \frac{3}{8}(a_{A}^{4}C_{40}^{A}+a_{B}^{4}C_{40}^{B})+\frac{9}{4}a_{A}^{2}a_{B}^{2}C_{20}^{A}C_{20}^{B}, \\
B_{2} &= -\frac{15}{2}(a_{A}^{4}C_{42}^{A}+a_{A}^{2}a_{B}^{2}C_{22}^{A}C_{20}^{B}), \;
B_{3} = 105a_{A}^{4}C_{44}^{A}, \\
B_{4} &= -\frac{15}{2}(a_{B}^{4}C_{42}^{B}+a_{A}^{2}a_{B}^{2}C_{22}^{B}C_{20}^{A}), \;
B_{5} = 105a_{B}^{4}C_{44}^{B}, \\
B_{6} &= \frac{9}{2}a_{A}^{2}a_{B}^{2}C_{22}^{A}C_{22}^{B}, \;
B_{7} = \frac{105}{2}a_{A}^{2}a_{B}^{2}C_{22}^{A}C_{22}^{B},
\end{aligned}
\end{equation*}
where the spherical harmonic coefficients of the primary asteroid $A$ are defined as \citep{balmino1994gravitational}
\begin{equation*}
    \begin{aligned}
        C_{20}^{A} &= \frac{1}{5a_{A}^{2}}\biggl(c_{A}^{2} - \frac{a_{A}^{2} + b_{A}^{2}}{2}\biggr),\; 
        C_{22}^{A} = \frac{1}{20a_{A}^{2}}\bigl(a_{A}^{2} - b_{A}^{2}\bigr),\\
        C_{40}^{A} &= \frac{15}{7}\Bigl[(C_{20}^{A})^{2} + 2(C_{22}^{A})^{2}\Bigr],\;C_{42}^{A} = \frac{5}{7} C_{20}^{A} C_{22}^{A},\\
        C_{44}^{A} &= \frac{5}{28}(C_{22}^{A})^{2}.
    \end{aligned}
\end{equation*}
The secondary asteroid $B$ holds the same definition of harmonic coefficients. 

To establish the Hamiltonian for describing spin--orbit coupling, a series of transformations should be performed \citep{lei2024perturbation}.

Initially, we introduce a set of canonical variables as follows:
\begin{equation}\label{Eq7}
\begin{aligned}
&r, \quad p_r = m_{\rm AB} \dot{r}, \\
&\psi_1 = \theta_A - \theta, \quad p_{\psi_1} = I_3^A \dot\theta_A, \\
&\psi_2 = \theta_B - \theta, \quad p_{\psi_2} =  I_3^B \dot\theta_B, \\
&\psi_3 = \theta, \quad p_{\psi_3} = m_{\rm AB} r^2 \dot\theta +I_3^A\dot\theta_A+I_3^B\dot\theta_B.
\end{aligned}
\end{equation}
Using the defined canonical variables, the Hamiltonian represented by Eq.~\eqref{Eq3} can be organized as
\begin{equation}\label{Eq8}
\begin{aligned}
\mathcal{H} =& -\frac{\mathcal{G} m_A m_B}{2a} + \frac{1}{2I_3^A} p_{\psi_1}^2 + \frac{1}{2I_3^B} p_{\psi_2}^2 \\
&-\frac{\mathcal{G} m_A m_B}{r^3} \Bigl[ A_1 + A_2 \cos 2\psi_1 + A_3 \cos 2\psi_2 \Bigr] \\
&-\frac{\mathcal{G} m_A m_B}{r^5} \biggl[ B_1 + B_2 \cos 2\psi_1 + B_3 \cos 4\psi_1 \\
& + B_4 \cos 2\psi_2 + B_5 \cos 4\psi_2\\
& + B_6 \cos (2\psi_1 - 2\psi_2)+B_7 \cos (2\psi_1 + 2\psi_2) \biggr].
\end{aligned}
\end{equation}
In this Hamiltonian, $\psi_3$ becomes a cyclic coordinate, thus its conjugate momentum $p_{\psi_3}$ is a motion integral. This conserved quantity is, in fact, the total angular momentum of the system, expressed by
\begin{equation}\label{Eq9}
    p_{\psi_3} = m_{\rm AB} r^2 \dot\theta +I_3^A\dot\theta_A+I_3^B\dot\theta_B = G_{{\rm tot}}.
\end{equation}
Now, we can see that the Hamiltonian given by Eq.~\eqref{Eq8} determines a 3 degrees of freedom (DOF) system, depending on the total angular momentum $G_{\rm tot}$. In order to study the spin--orbit coupling from perturbation theory, it is necessary to perform elliptic expansions as follows \citep{hughes1981computation}:
\begin{equation}\label{Eq10}
\left( \frac{a}{r} \right)^l \cos(mf) = \sum_{s=-\infty}^{\infty} X_s^{-l,m}(e) \cos(sM),\\
\end{equation}
where $M$ is the mean anomaly, and $X_s^{-l,m}(e)$ is the Hansen coefficient, which can be expressed as a power series in eccentricity \citep{murray1999solar}. 

Furthermore, it is convenient to introduce the true anomaly $\lambda = \varpi + M$. Then, the set of Delaunay variables can be introduced as follows:
\begin{equation}\label{Eq11}
\begin{aligned}
l      &= \lambda - \varpi,       &\quad L      &= m_{\rm AB} \sqrt{\mu a},\\
\gamma_1 &= \theta_A - \varpi,     &\quad \Gamma_1 &=I_3^A\dot\theta_A,\\
\gamma_2 &= \theta_B - \varpi,     &\quad \Gamma_2 &= I_3^B\dot\theta_B,\\
\gamma_3 &= \varpi,                &\quad \Gamma_3 &= G_{\rm tot},
\end{aligned}
\end{equation}
where $\mu={\cal G}(m_A+m_B)$ is the gravitational constant. For convenience of computation, we normalize the time and space variables in the following system of units:
\begin{equation}\label{Eq12}
[L] = a_A,\quad [M] = m_{\rm AB},\quad [T] = \sqrt{\frac{a_A^3}{\mu}},
\end{equation}
where $[L]$, $[M]$, and $[T]$ represent the units of length, mass, and time, respectively.

Under the normalized units, the 3-DOF Hamiltonian can be expressed in Delaunay variables as follows:
\begin{equation}\label{Eq13}
\begin{aligned}
\mathcal{H} = & -\frac{1}{2L^2} + \frac{\Gamma_1^2}{2I_3^A} + \frac{\Gamma_2^2}{2I_3^B} - \sum_{n=-\infty}^{\infty} \left[c_{n,0} \cos(n l)\right. \\
&+c_{n,1} \cos(n l - 2 \gamma_1) + c_{n,2} \cos(n l - 2 \gamma_2) \\
&+c_{n,3} \cos(n l - 4 \gamma_1) + c_{n,4} \cos(n l - 4 \gamma_2) \\
&+ c_{n,5} \cos(n l + 2 \gamma_1 - 2 \gamma_2)\\
&+ \left.c_{n,6} \cos(n l - 2 \gamma_1 - 2 \gamma_2)\right],
\end{aligned}
\end{equation}
where the coefficients are given by
\begin{equation*}
\begin{aligned}
c_{n,0}&=\frac{A_1}{L^6} X_n^{-3,0}(e) + \frac{B_1}{L^{10}} X_n^{-5,0}(e),\;c_{n,3}=\frac{B_3}{L^{10}} X_n^{-5,4}(e),\\
c_{n,1}&=\frac{A_2}{L^6} X_n^{-3,2}(e) + \frac{B_2}{L^{10}} X_n^{-5,2}(e),\;c_{n,4}=\frac{B_5}{L^{10}} X_n^{-5,4}(e),\\
c_{n,2}&=\frac{A_3}{L^6} X_n^{-3,2}(e) + \frac{B_4}{L^{10}} X_n^{-5,2}(e),\;c_{n,5}=\frac{B_6}{L^{10}} X_n^{-5,0}(e),\\
c_{n,6}&=\frac{B_7}{L^{10}} X_n^{-5,4}(e).
\end{aligned}
\end{equation*}
It can be observed that the set of canonical variables $(l,L)$ is used to describe the evolution of mutual orbit, the set of $(\gamma_1,\Gamma_1)$ is used to describe the rotation evolution of the primary asteroid and the set of $(\gamma_2,\Gamma_2)$ is for the rotation evolution of the secondary asteroid.

The Hamiltonian canonical relations lead to the equations of motion, given by
\begin{equation}\label{Eq14}
\begin{aligned}
\frac{{\rm d}l}{{\rm d}t} &= \frac{\partial \mathcal{H}}{\partial L}, &
\frac{{\rm d}L}{{\rm d}t} &= -\frac{\partial \mathcal{H}}{\partial l}, \\
\frac{{\rm d}\gamma_1}{{\rm d}t} &= \frac{\partial \mathcal{H}}{\partial \Gamma_1}, &
\frac{{\rm d}\Gamma_1}{{\rm d}t} &= -\frac{\partial \mathcal{H}}{\partial \gamma_1}, \\
\frac{{\rm d}\gamma_2}{{\rm d}t} &= \frac{\partial \mathcal{H}}{\partial \Gamma_2}, &
\frac{{\rm d}\Gamma_2}{{\rm d}t} &= -\frac{\partial \mathcal{H}}{\partial \gamma_2}.
\end{aligned}
\end{equation}
By numerically integrating the equations of motion, we can obtain the time evolution of spin and orbit variables of the binary asteroid system. Accordingly, it is possible for us to construct the resonant network of spin--orbit coupling, which will be explored in the coming sections.

\section{Resonant networks}
\label{Sect3}

In the classical spin--orbit resonant model, the resonance width of the synchronous spin--orbit resonance can be measured by the asteroid's shape parameter \citep{murray1999solar}:
\begin{equation}
\begin{aligned}
d_{1:1}=\alpha_i=\sqrt{3\frac{a_i^2-b_i^2}{a_i^2+b_i^2}},\; (i=A,B),
\end{aligned}
\end{equation}
where $a_i$ and $b_i$ are the semimajor and semi-intermediate axes of the asteroid. In this work, we also adopt $\alpha_i$ as the shape parameter. It quantifies the nonsphericity of the asteroids while simultaneously providing a priori reference for the resonance structures within the network. It is important to note that, owing to spin--orbit coupling and the dimensionless formulation used here, $\alpha_i$ cannot precisely describe the actual width in the spin--orbit coupling problem.

\subsection{Parameter Setting}
\label{sec3.1}

To visualize the global structure of resonances in binary asteroids, we construct a resonant network by integrating the set of canonical equations given by Eq. \eqref{Eq14}. We define a two-dimensional parameter space spanned by $k_1=\frac{\dot\gamma_{10}}{n_0}$ and $k_2=\frac{\dot\gamma_{20}}{n_0}$, representing the ratios of the initial spin rates of the primary and secondary asteroids to the initial orbital mean motion. In practice, the two-dimensional parameter space is discretized into an $n\times n $ grid of square cells with step of $h$. Each point of the grid corresponds to a distinct initial condition of the binary system.

A key feature of our setup is that all grid points are assigned with the same initial orbital angular momentum $L_0$, resulting in the same initial mean motion $n_0=\frac{1}{L^3_0}$ \footnote{This setup employs the unperturbed Hamiltonian, allowing us to uniquely determine the initial mean motion $n_0$ via $n_0=\dot l_0=\frac{1}{L_0^3}$, facilitating a systematic scanning of the phase space.}. It should be noted that the global structure is insensitive to the choice of $L_0$. Together with the prescribed initial spin angles $(\theta_{10}, \theta_{20})$ and mean anomalies $l_0$, the complete initial condition at any grid point can be fully determined by ($k_1,k_2$).

\subsection{The Second-Derivative-Based Index}
\label{sec3.2}

To characterize the structure within a parameter space, it is crucial to select an appropriate dynamical indicator. In this regard, \citet{daquin2023detection} have shown that zeroth-order measures, like the maximum variation in semi-major axis $\delta a$, may underestimate the resonant and chaotic architecture and fail in recognizing thinner secondary structures. To overcome this problem, they defined the second-derivative-based indicator $\Vert\Delta D \Vert$, which can be used to capture chaos and reconstruct resonant structures precisely. 

In this work, we adopt the normalized second-derivative metric to study the resonant network of spin--orbit coupling. For convenience, we normalize the  second-derivative indicator $\Vert \Delta  D \Vert$ as
\begin{equation}\label{Eq16}
\Vert\Delta  D \Vert  =\frac{\vert \Delta \, \delta a \vert }{ \delta a} +\frac{\vert \Delta \, \delta \Gamma_1 \vert }{ \delta \Gamma_1} +\frac{\vert \Delta \, \delta \Gamma_2 \vert }{ \delta \Gamma_2} ,
\end{equation}
where $\delta f \;(f=a,\Gamma_1,\Gamma_2)$ denotes the maximum variation of $f$ over the considered period of time and the second-order derivative $|\Delta \,\delta f|$ is approximated by a central finite-difference scheme \citep{daquin2023detection}
\begin{equation*}
\begin{aligned}
\vert\Delta \,\delta f \vert & \approx\frac{\vert\delta f(x_{i+1},y_i)-2\delta f(x_i,y_i)+\delta f(x_{i-1},y_i)\vert}{h^2}\\
&+\frac{\vert\delta f(x_i,y_{i+1})-2\delta f(x_i,y_i)+\delta f(x_i,y_{i-1})\vert}{h^2}.
\end{aligned}
\end{equation*}
Under spin--orbit coupling, the system dynamics involve the exchange of spin and orbital angular momentum. The indicator $\Vert \Delta D \Vert$ is thus designed to combine both changes. In particular, $\vert\Delta \delta a \vert$ is used to specify the variation of orbit, and $\vert\Delta \delta \Gamma_1 \vert$ and $\vert\Delta \delta \Gamma_2 \vert$ are used to specify the spin variations of both asteroids. The second-derivative-based index has been successfully applied to reveal dynamical structures of high-order and secondary resonances in the spin--orbit problem \citep{lei2024dynamical}.
    
\subsection{Dynamical Structures}
\label{sec3.3}

The second-derivative-based index is adopted to reveal the dynamical structure of binary asteroid systems. In particular, the angles are initially assumed at $l_0=\gamma_{10}=\gamma_{20}=0$ and the eccentricity at $e_0=0.05$. In addition, the initial orbital angular momentum is fixed at $L_0=2.5$ (i.e., $a \approx 6.25\,a_A$), with $m_A=m_B$. We produce the distribution of the second-derivative-based indicator in the parameter space of $(k_1,k_2)$ for binary asteroid systems with different combinations of $\alpha_A$ and $\alpha_B$.

\begin{figure*}
\centering
\includegraphics[width=0.8\linewidth]{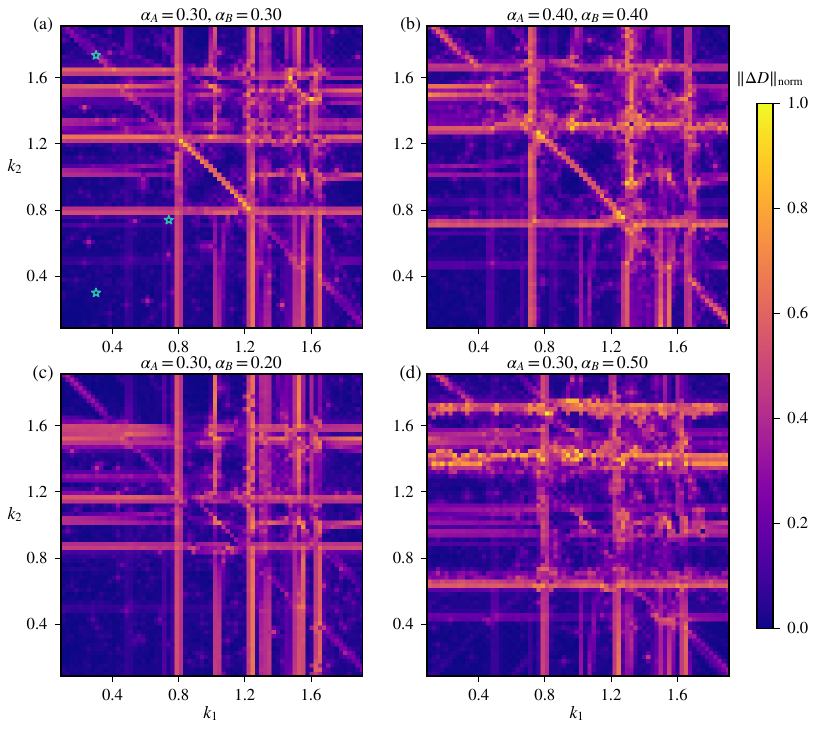}
\caption{Resonant networks of spin--orbit coupling in the two-dimensional parameter space of ($k_1 = \dot{\gamma}_{10} / n_0, k_2 = \dot{\gamma}_{20} / n_0$) for binary asteroid systems with four different combinations of shape parameters $(\alpha_A,\alpha_B)$. For each case, $ 80 \times 80$ grids of initial conditions of rotation velocity (with identical initial orbital angular momentum at $L_0=2.5$) are considered for numerical simulation over 100 orbital periods. In the top left panel, three special points on the spin--orbit--spin resonance stripe ($k_1$ = 0.300, $k_2$ = 1.736) and on the spin--spin resonance stripe ($k_1$ = 0.3, $k_2$= 0.3; $k_1$=0.739, $k_2$=0.741) are marked by blue stars and their associated trajectories are presented in Fig.~\ref{Fig3}.}
\label{Fig2}
\end{figure*}

\begin{figure}
    \centering
    \begin{minipage}[b]{0.48\textwidth}
        \centering
        \includegraphics[width=\linewidth]{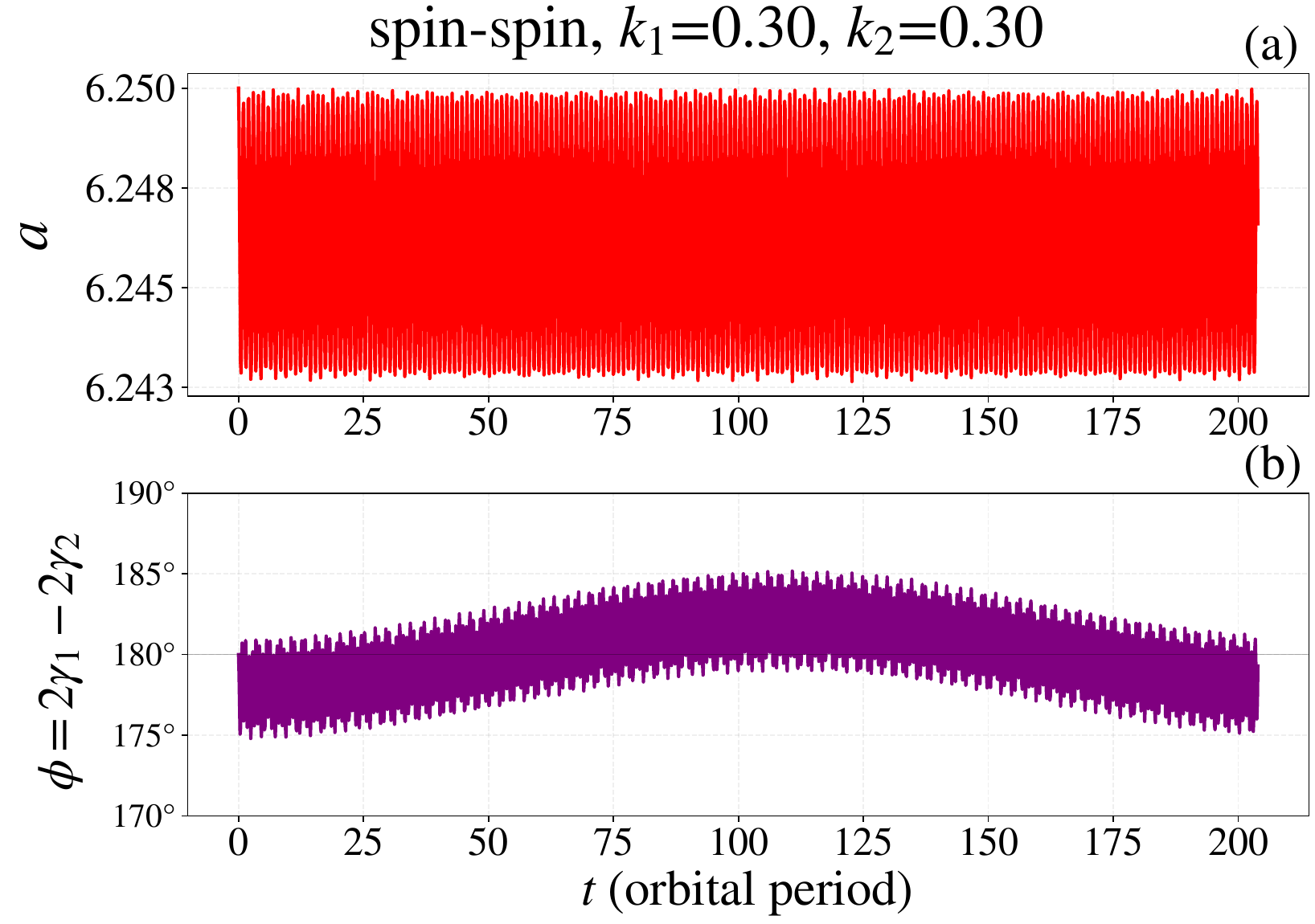}
    \end{minipage}
    \hfill
     \begin{minipage}[b]{0.48\textwidth}
        \centering
        \includegraphics[width=\linewidth]{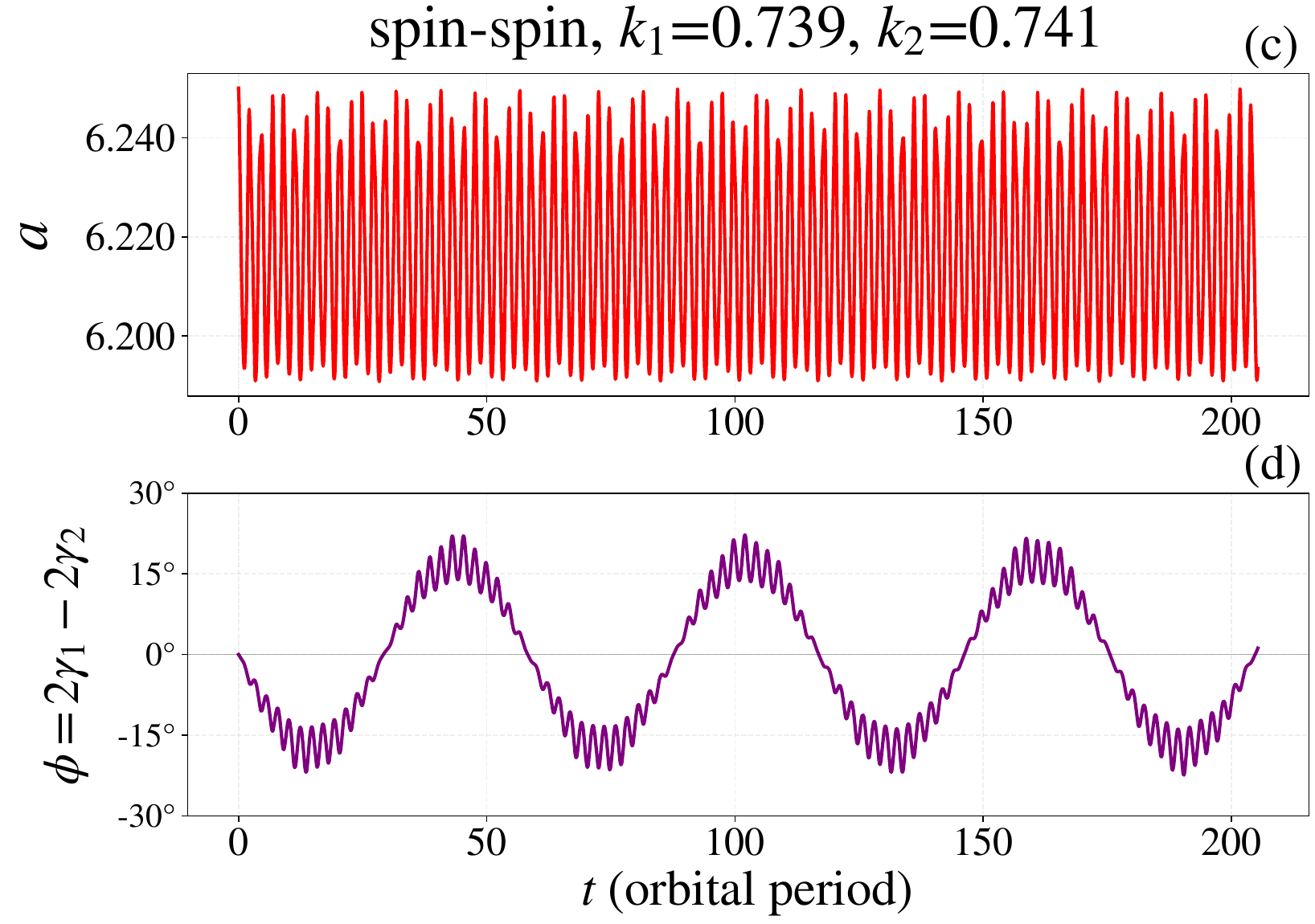}
    \end{minipage}
    \hfill
     \begin{minipage}[b]{0.48\textwidth} 
        \centering
        \includegraphics[width=\linewidth]{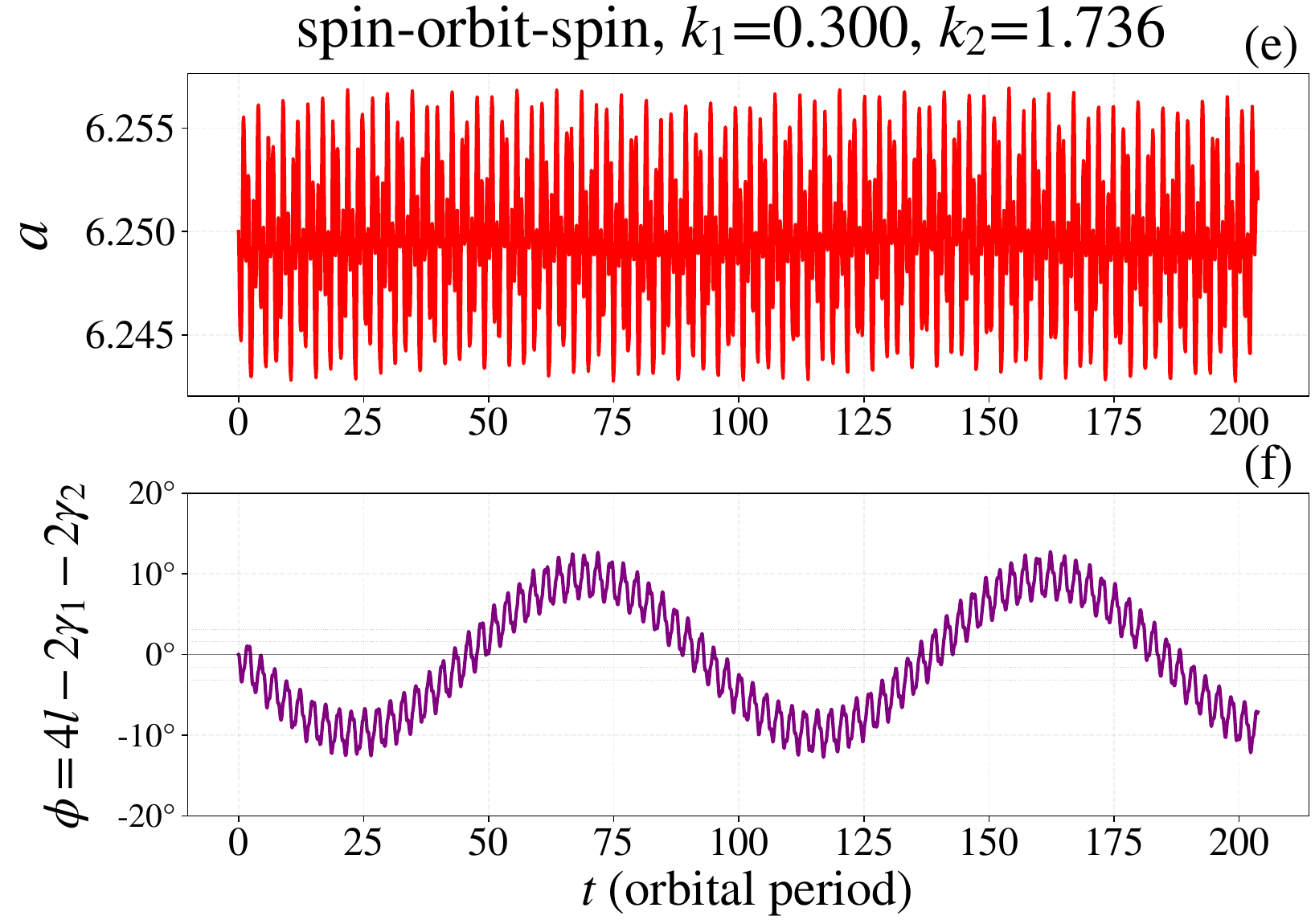}
    \end{minipage} 
   \caption{Time evolution of semimajor axis and critical arguments for the resonant trajectories in the model of $\alpha_A = \alpha_B = 0.3$, which are marked by blue stars in the top left panel of Fig.~\ref{Fig2}. The top two panels are for the spin--spin resonance under the condition of $k_1=k_2 = 0.3$. The middle two panels are for the spin--spin resonance under the condition of $k_1=0.739,\;k_2 = 0.741$. The bottom two panels are for the spin--orbit--spin resonance under the condition of $k_1 = 0.300$ and $k_2 = 1.736$. The orbital period is given by $T=2\pi L_0^3$ where $L_0=2.5$ is the initial orbital angular momentum at zero eccentricity.}
    \label{Fig3}
\end{figure}

Fig.~\ref{Fig2} presents the complex structures for four representative binary asteroid systems. For each case, $80\times80$ grids of initial conditions are considered for numerical simulation over $\sim$$100$ orbital periods. The choice of integration period is a critical issue. In the spin--orbit coupling problem, \citet{jafari2023surfing} have shown that  selecting 100 orbital periods is appropriate for computing fast Lyapunov index (FLI) to map the phase-space structures. Supplementary resonant networks for other physical configurations (e.g., mass ratios and shapes) can be found in Appendix \ref{A2}. 

It is observed from Fig.~\ref{Fig2} that (a) there is a network of spin--orbit resonances in the plane of $(k_1,k_2)$, such as 1:1, 3:2, and 1:2 resonances; (b) in each case there is a special stripe structure with a slope of $\sim -1$, extending from the left top corner to the bottom right corner (the dynamics of such a stripe-like structure are studied in Sect.~\ref{Sect4}); (c) the resonance width of the 1:1 resonance is positively correlated with the shape parameter $\alpha_A$ or $\alpha_B$, which is in agreement with the known conclusion \citep{murray1999solar}; (d) the overlapping zone of nearby resonances may appear when the shape parameter increases, leading to chaotic evolutions. 

Notably, there is a special region in the parameter space where the primary and secondary asteroids' synchronous spin--orbit resonances coexist. We refer to this region as the doubly synchronous region (please see the central region of each panel). Inside this region,  the dynamics of binary asteroids are governed by synchronous resonances of both asteroids. In the case of $\alpha_A = \alpha_B$, the doubly synchronous region corresponds to a square and, in the case of $\alpha_A \ne \alpha_B$, it corresponds to a rectangle. It is not difficult to understand this phenomenon. In particular, the stripe structure is bright inside the doubly synchronous region for the case of $\alpha_A = \alpha_B$. However, the stripe becomes faint in the case of $\alpha_A \ne \alpha_B$. This indicates that, as the two synchronous spin--orbit resonances become comparable in strength, they couple together and give rise to a new, prominent secondary resonance inside the doubly synchronous region. 

Outside the synchronous spin--orbit resonance region, a stripe with a slope of $\sim$$-1$ can also be observed. It is potentially associated with the spin--orbit--spin resonance. Its brightness is weaker than the stripes inside the doubly synchronous region, meaning that it is caused by a higher-order resonance. In addition, outside the synchronous resonance region we can see a faint stripe with a slope of $\sim$$+1$, specifically in the lower-left corner of Fig.~\ref{Fig2}(d). It may be associated with the spin--spin resonance. 

Starting from the initial conditions on the spin--spin and spin--orbit--spin resonant stripes shown in Fig.~\ref{Fig2}(a), we numerically propagate the spin--orbit coupling dynamical model of binary asteroids. Fig.~\ref{Fig3} shows the resulting time histories of the semimajor axis and the associated resonant argument $\phi$, illustrating the dynamical behaviors of these resonances. We can see that the spin--spin resonance has no influence upon the evolution of semimajor axis, while the spin--orbit--spin resonance can lead to periodic oscillations of the semimajor axis. The resonant argument $\phi$ is librating around either $180^{\circ}$ or  $0^{\circ}$ for the spin--spin resonance (see panels b and d), and it librates around $0^{\circ}$ for the spin--orbit--spin resonance (see panel f). This point can be understood with the help of analytical study performed in Sect.~\ref{Sect4}. 

One remark is made here. The two examples of spin--spin resonance demonstrate that this resonance is stably present in the parameter space of Fig.~\ref{Fig2}(a). Due to its high-order character, the corresponding stripe with a slope of $\sim$$+1$ is extremely faint and difficult to observe directly.

\section{Dynamics in different regions}
\label{Sect4}

\begin{figure*}
    \centering
    \begin{minipage}{0.4\textwidth}
        \centering
        \includegraphics[width=\linewidth]{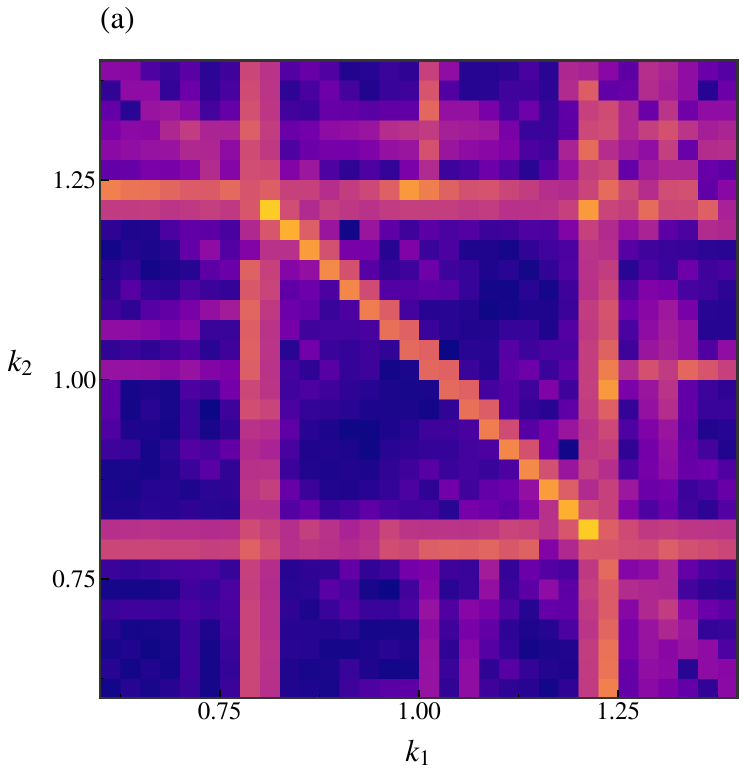}
    \end{minipage}
    \hspace{0.5cm} 
    \begin{minipage}{0.4\textwidth}
        \centering
        \includegraphics[width=\linewidth]{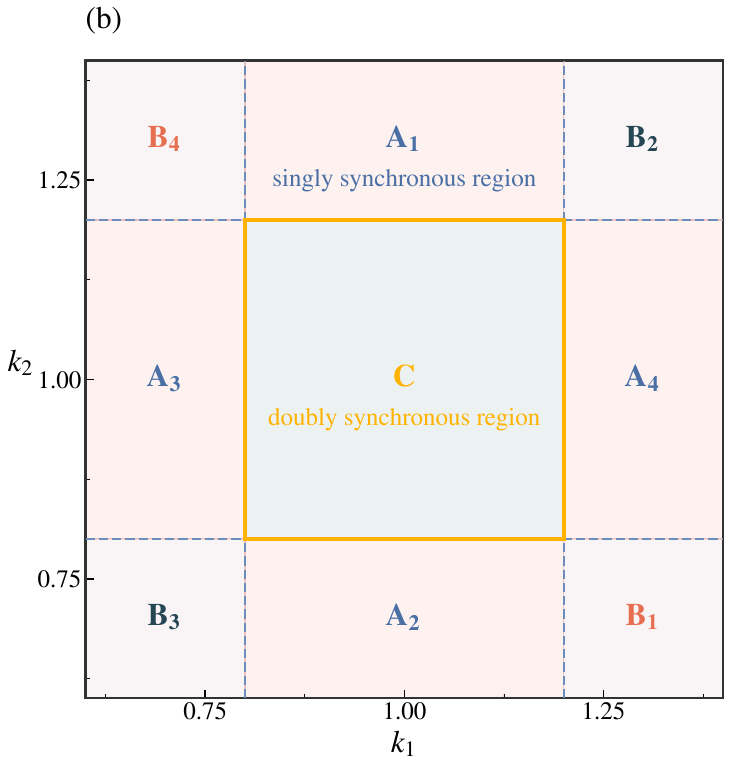}
    \end{minipage}
    \caption{Resonant network in the interested region (\textit{left panel}) and preliminary partition of the synchronous region (\textit{right panel}) under the model of $\alpha_A = \alpha_B = 0.3$. This region is further divided into three subregions, denoted by $A$ (singly synchronous regions), $B$ (nonsynchronous regions), and $C$ (doubly synchronous region).}
    \label{Fig4}
\end{figure*}

The structure arising in the resonant network (see Fig.~\ref{Fig2}) provides us a global picture about the dynamics of spin--orbit coupling in binary asteroid systems. In this work, we are interested in the dynamics in the vicinity of synchronous resonances of both asteroids. Thus, we concentrate on a representative region near the synchronous resonances (see the left panel of Fig.~\ref{Fig4}). This region encompasses singly synchronous resonance, spin--spin and spin--orbit--spin resonances, and doubly synchronous resonances. 

Based on the observation of the resonant network structure discussed in Sect.~\ref{Sect3}, we can divide the region of interest into three distinct domains according to the existence of synchronous resonance (see the right panel of Fig.~\ref{Fig4}).
\begin{itemize}
\item Domain $A$ including $A_i$ ($i=1,2,3,4$), corresponding to singly synchronous zones, is dominated by singly synchronous spin--orbit resonance.
\item Domain $B$ including $B_i$ ($i=1,2,3,4$), corresponding to non-synchronous zones, is dominated by spin--orbit--spin and spin--spin resonances.
\item Domain $C$, corresponding to doubly synchronous zones, is dominated by synchronous resonances of both asteroids.
\end{itemize}

This simulation-derived preclassification naturally prompts two theoretical inquiries: First, the resonance boundaries suggested by numerical metrics should be analytically delineated, thereby allowing a rigorous partition of each resonance's domain.
Second, the distinct dynamics within these partitions require characterization, which involves the following: for domains $A$ and $B$, identifying the governing perturbative terms and their associated resonant structures; for domain $C$, determining the conditions for the newly identified secondary resonance and elucidating the coupling of two parent synchronous spin--orbit resonances.

Guided by these inquiries, our subsequent analysis of regions $A$, $B$, and $C$ seeks to uncover their governing dynamics, thereby refining the preclassification.

\begin{figure*}
    \centering
    \begin{minipage}{0.31\textwidth}
        \centering
        \includegraphics[width=\linewidth]{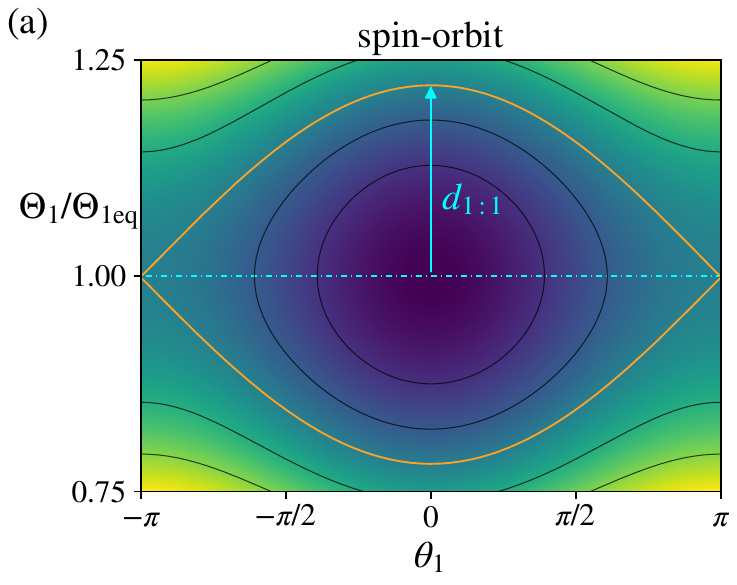}
    \end{minipage}
    \hspace{0.1cm} 
    \begin{minipage}{0.31\textwidth}
        \centering
        \includegraphics[width=\linewidth]{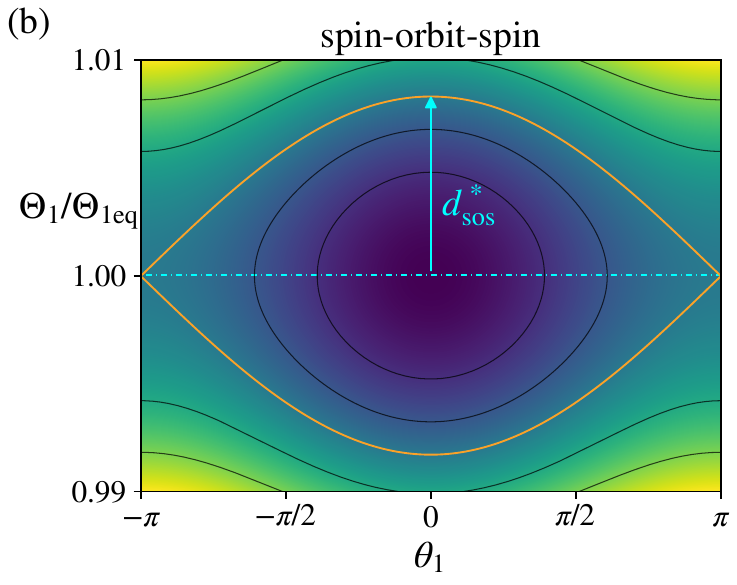}
    \end{minipage}
    \hspace{0.1cm}
    \begin{minipage}{0.31\textwidth}
        \centering
        \includegraphics[width=\linewidth]{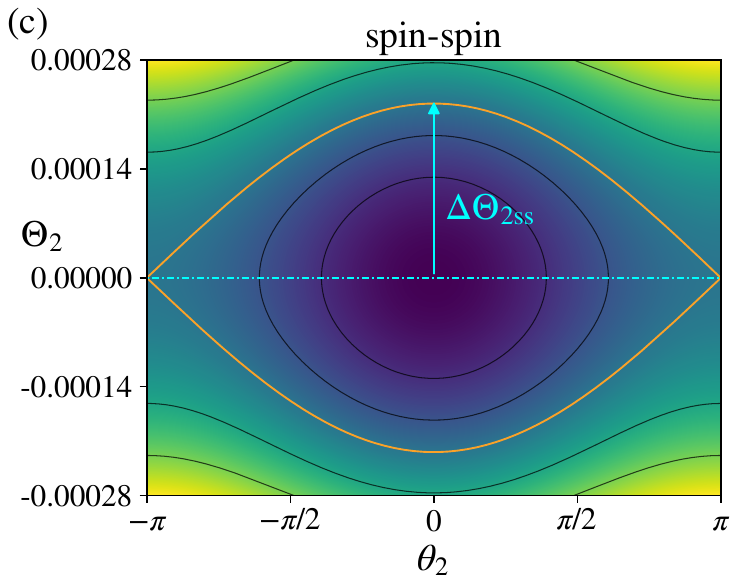}
    \end{minipage}
    \caption{Phase portraits of the spin--orbit synchronous resonance (a), spin--orbit--spin resonance (b) and spin--spin resonance (c) in the binary asteroid system with $\alpha_A=\alpha_B=0.3$. The half-width of each resonance is indicated by an arrow. The motion integral is \( I = L + \Gamma_1 = 2.55 \) in panel (a), \( I = L + \Gamma_1 + \Gamma_2 = 2.60 \) in panel (b), and \( I = L + \Gamma_1 + \Gamma_2 = 2.57 \) in panel (c).}
    \label{Fig5}
\end{figure*}

\subsection{Dynamics inside Region $A$}
\label{sec4.1}

The synchronous spin--orbit resonance, as the most prominent zeroth order (in eccentricity) resonance in binary asteroid systems, dominates the dynamics of region $A$. Its properties under the coupled model have been extensively investigated \citep{celletti1990analysis1,celletti1990analysis2,celletti2000hamiltonian,hou2017note,lei2024primary}. For the purpose of self-consistency, in this subsection we revisit this problem and derive an explicit expression for its resonance width, in order to understand the numerical results. 

We first derive the resonant Hamiltonian through averaging within the spin--orbit coupling framework. Subsequently, we construct its phase portrait to characterize the equilibrium points and separatrix, and apply the pendulum approximation to derive an explicit expression for the resonance half-width.

Taking the resonance of the primary asteroid as an example (the analysis for the secondary is the same), we average out those short-period perturbations and neglect those fourth-order influences associated with $\frac{A_1}{L^6}$ and $\frac{B_2}{L^{10}}$ \citep{lei2024primary}. Further truncating the Hamiltonian at the first order in eccentricity, we can obtain the resonant Hamiltonian as follows:
\begin{equation}
    \begin{aligned}
    \mathcal{H}_{1:1}=-\frac{1}{2L^2}+\frac{\Gamma_1^2}{2I^A_3}-\frac{A_2}{L^6}\cos(2l-2\gamma_1).
    \end{aligned}
    \label{Eq.ch4.1-Hamiltonian-spin--orbit}
\end{equation}\\
For convenience, we introduce the following canonical transformation:
\begin{equation}
\begin{aligned}
    \Theta_1&=-\frac{1}{2}\Gamma_1,\quad \theta_1=2l-2\gamma_1,\\
    I&=\Gamma_1+L,\quad \psi=l,
\end{aligned}
\label{Eq.ch4.1-canonical-transformation-spin--orbit}
\end{equation}
where $\theta_1$ is the critical angle of the synchronous spin--orbit resonance. Applying this canonical transformation to Eq. \eqref{Eq.ch4.1-Hamiltonian-spin--orbit}, we obtain the resulting 1-DOF Hamiltonian, which characterizes the dynamics of synchronous spin--orbit resonance:
\begin{equation}
 \mathcal{K}_{1:1}={\cal K}_0+V\cos\theta_1,
\label{Eq.ch4.1-Kamiltonian-spin--orbit}
\end{equation}
where
\begin{equation*}
{\cal K}_0 = -\frac{1}{2(2\Theta_1+I)^2}+\frac{2\Theta_1^2}{I^A_3},
\end{equation*}
and 
\begin{equation*}
V=-\frac{A_2}{(2\Theta_1+I)^6}.
\end{equation*}
In the resonant Hamiltonian model, the angular coordinate $\psi$ is a cyclic variable, showing that its conjugate momentum $I=\Gamma_1 + L$ is an integral of motion. It indicates that, during the spin--orbit synchronous resonance, there is an exchange between the rotational and orbital angular momentum. 

Based on the resonant Hamiltonian $\mathcal{K}_{1:1}$, the phase portrait for the synchronous resonance is plotted in the space of $(\theta_1,\Theta_1)$. With $I=2.55$, the portrait shown in panel (a) of Fig.~\ref{Fig5} reveals the standard cat-eye structure: a stable center at $\theta_1 = 0$, bounded by a separatrix (yellow curve) passing through the saddle point at $\theta_1 = \pi$.

In accordance with \citet{morbidelli2002modern}, we expand $\mathcal{K}_{0}$ around the nominal location of synchronous spin--orbit resonance in a second-order Taylor series (the so-called pendulum approximation) as follows:
\begin{equation}
    \begin{aligned}
     \frac{\partial \mathcal{K}_0}{\partial \Theta_1}&=\frac{2}{(2\Theta_1+I)^3}+\frac{4\Theta_1}{I^A_3},\\
    \frac{\partial^2 \mathcal{K}_0}{\partial \Theta_1^2}&=-\frac{12}{(2\Theta_1+I)^4}+\frac{4}{I^A_3}.
    \end{aligned}
    \label{Eq.K11}
\end{equation}
At the equilibrium point, it holds $\frac{\partial K_0}{\partial \Theta_1}=0$, yielding that $\Theta_{1{\rm eq}}=-\frac{I^A_3}{2L_{\rm eq}^3}$, where $L_{\rm eq}$ denotes the value of the orbital angular momentum at the equilibrium point.

Under the pendulum approximation, the resonant Hamiltonian for synchronous spin--orbit resonance can be approximated as
\begin{equation}
 \Delta \mathcal{K}_{1:1}=\frac{1}{2}M(\delta\Theta_1)^2+ V \cos\theta_1,
\label{Eq.K11}
\end{equation}
where 
\begin{equation*}
M=\frac{\partial^2 {\mathcal{K}}_0}{\partial \Theta_1^2}|_{\Theta_1=\Theta_{1{\rm eq}}}=-\frac{12}{L_{{\rm eq}}^4}+\frac{4}{I^A_3}.
\end{equation*}
Here, $\Delta \mathcal{K}_{1:1}$ is just a pendulum approximation of the resonant Hamiltonian, from which we can derive the expression of the half-width of the synchronous spin--orbit resonance in the following form:
\begin{equation}
    \begin{aligned}
    d_{1:1}=\frac{\Delta \dot \gamma_1}{n_0}=\frac{1}{n_0}\frac{2\Delta \Theta_1}{I^A_3}=\frac{4L_{{\rm 0}}^3}{I^A_3}\sqrt{\vert\frac{V}{M}\vert}. 
    \end{aligned}
    \label{Eq.ch4.1-d-spin--orbit}
\end{equation}
The half-width of spin--orbit synchronous resonance is marked by a blue arrow in Fig.~\ref{Fig5} (a).

The Hamiltonian given by Eq.~\eqref{Eq.K11} shows that the location of resonant center is determined by the signs of $M$ and $V$. It is noteworthy that, in our model the condition $V<0$ always holds, and thus the location of the synchronous spin--orbit resonance center is uniquely determined by the sign of the coefficient $M$. In particular, if $M>0$, the resonant center is located at $\theta_{1}=0$, whereas if $M<0$ it shifts to $\theta_{1}=\pi$. This analytical criterion is totally equivalent to the factor $S_{1}$ and $S_{1:1}$ discussed in \cite{hou2017note} and \cite{lei2024primary}.

\subsection{Dynamics inside Region $B$}
\label{sec4.2}

In this subsection, we analyze high-order resonances occupying region $B$ (the nonsynchronous resonance zone), including the spin--orbit--spin and spin--spin resonances. For such two types of resonances, the conventional averaging method (i.e., the lowest-order perturbative treatment) may introduce significant deviations \citep{hou2017note,lei2024perturbation}. To address this problem, we employ Lie-series transformation to derive high-order resonant Hamiltonian. Regarding this point, similar discussions  from a different perspective can be found in \citet{hou2017note}.

In practice, Hori's iterative approach is taken to perform Lie-series transformation \citep{hori1966theory}: 
\begin{equation}\label{Eq.ch4.2-Hori}
    \begin{aligned}
    \mathcal{K}_0=& \mathcal{H}_0, \\
     \mathcal{K}_1=& \mathcal{H}_1+\{\mathcal{H}_0 , W_1\},\\
    \mathcal{K}_2=&\mathcal{H}_2+\{\mathcal{H}_1 , W_1\}+\frac{1}{2!}\{ \{\mathcal{H}_0 , W_1\},W_1 \} + \{\mathcal{H}_0 , W_2\}.
    \end{aligned}
\end{equation}
The original Hamiltonian $\cal{H}$ is transformed into a resonant Hamiltonian $\cal{K}$ via the generating function $W = W_1 + W_2$.

To reduce the degrees of freedom of $\cal{K}$ for further analysis, we need to introduce a canonical transformation which is convenient for spin--spin and spin--orbit--spin resonances as follows:
\begin{equation}
    \begin{aligned}
    \Theta_1&=-\frac{1}{4}(\Gamma_1+\Gamma_2),\quad \theta_1=4l-2\gamma_1-2\gamma_2,\\
    \Theta_2&=\frac{1}{4}(\Gamma_1-\Gamma_2),\quad \theta_2=2\gamma_1-2\gamma_2, \\
    I&=\Gamma_1+\Gamma_2+L, \quad\psi=l.
    \end{aligned}
    \label{eq.ch4.2-canonical-transformation-spin--orbit--spin}
\end{equation}
Under such a new set of variables, the pair of $(\theta_1,\Theta_1)$ corresponds to the spin--orbit--spin resonance and the pair of $(\theta_2,\Theta_2 )$ corresponds to the spin--spin resonance. $\theta_1$ and $\theta_2$ are their resonant angles. In the following subsections, we apply this analytical framework to each resonance.

\subsubsection{Spin--Orbit--Spin Resonance}
\label{sec4.2.1}

Truncating at the first order of eccentricity in Eq. \eqref{Eq13}, we can write the Hamiltonian as follows:
\begin{equation}\label{Eq.ch4.2-Hamiltonian-spin--orbit--spin}
\mathcal{H}=\mathcal{H}_0+\mathcal{H}_1+\mathcal{H}_2,
\end{equation}
where
\begin{equation*}
\begin{aligned}
\mathcal{H}_0=&-\frac{1}{2L^2}+\frac{\Gamma_1^2}{2I^A_3}+\frac{\Gamma_2^2}{2I^B_3},\\
\mathcal{H}_1=&-\frac{A_1}{L^6}-\frac{3A_1}{L^6}e\cos(l)\\
&+\frac{A_2}{2L^6}e\cos(l-2\gamma_1)-\frac{A_2}{L^6}\cos(2l-2\gamma_1)\\
&-\frac{7A_2}{2L^6}e\cos(3l-2\gamma_1)
+\frac{A_3}{2L^6}e\cos(l-2\gamma_2)\\
&-\frac{A_3}{L^6}\cos(2l-2\gamma_2)-\frac{7A_3}{2L^6}e\cos(3l-2\gamma_2),
\end{aligned}
\end{equation*}
and
\begin{equation*}
\begin{aligned}
    \mathcal{H}_2=-\frac{B_7}{L^{10}}  \cos( 4l -2 \gamma_1 - 2 \gamma_2).
\end{aligned}
\end{equation*}
In particular, the eccentricity should be expressed as a function of the canonical variables in the manner,
\begin{equation*}
e=\sqrt{1-\left(\frac{G_{\rm tot}-\Gamma_1-\Gamma_2}{L}\right)^2},
\end{equation*}
and its partial derivatives can be obtained as
\begin{equation*}
\frac{\partial e}{\partial L}=\frac{1-e^2}{Le},\quad \frac{\partial e}{\partial \Gamma_1}=\frac{\partial e}{\partial \Gamma_2}=\frac{\sqrt{1-e^2}}{Le}.
\end{equation*}
$\mathcal{H}_2$ is the perturbative term in $\mathcal{H}$, which is directly associated with the spin--orbit--spin resonance. As noted before, however, the perturbation in $\mathcal{H}_1$ cannot be averaged out directly, because it has second-order contribution to the considered resonance. To this end, we formulate a new Hamiltonian $\mathcal{K}$ via the Lie-series transformation by introducing the second-order contribution from $\mathcal{H}_1$. The resulting $\mathcal{K}_1$ satisfies
\begin{equation}
    \begin{aligned}
      \mathcal{K}_1=\mathcal{H}_1+\{\mathcal{H}_0 , W_1\}=-\frac{A_1}{L^6} 
    \end{aligned}
    \label{Eq.ch4.2-K1-equation}
\end{equation}
By solving this equation, we can obtain the expression of generating function $W_1$, which is shown by Eq.~\eqref{W1} in Appendix \ref{A1}.

Substituting the expression of $W_1$ into Eq. \eqref{Eq.ch4.2-Hori}, averaging out short-period terms and applying the canonical transformation \eqref{eq.ch4.2-canonical-transformation-spin--orbit--spin}, we could derive the expression for $\mathcal{K}_{{\rm sos}}$ applicable to study spin--orbit--spin resonance as follows: 
\begin{equation*}
\mathcal{K}_{\rm sos}=\mathcal{K}_0+\mathcal{K}_1+V_{\rm sos}\cos \theta_1,
\end{equation*}
where 
\begin{equation*}
\mathcal{K}_0=-\frac{1}{2(4\Theta_1+I)^2}+\frac{2(\Theta_1-\Theta_2)^2}{I^A_3}+\frac{2(\Theta_1+\Theta_2)^2}{I^B_3},
\end{equation*}
and
\begin{equation*}
\mathcal{K}_1=-\frac{A_1}{(4\Theta_1+I)^6}.
\end{equation*}
The expression of $V_{\rm sos}$ is provided by Eq. \eqref{Appendix V s-o-s} in Appendix \ref{A1} .

The subsequent discussion is similar to that of the spin--orbit synchronous resonance in Sect.~\ref{sec4.1}. Noting that the variables corresponding to the spin--orbit--spin resonance are  $\Theta_1$ and $\theta_1$, and expanding $\mathcal{K}_{\rm sos}$ at the nominal position of spin--orbit--spin resonance, we obtain the pendulum approximation for the spin--orbit--spin resonance as follows:
\begin{equation}
 \Delta \mathcal{K}_{\rm sos}=\frac{1}{2}M(\delta\Theta_1)^2+V_{\rm sos}\cos\theta_1,
\label{Eq.ch4.2-d-spin--orbit--spin}
\end{equation}
where
\begin{equation*}
 M_{\rm sos}=\frac{\partial^2 ({\mathcal{K}}_0+{\mathcal{K}}_1)}{\partial \Theta_1^2}|_{\Theta_1=\Theta_{1{\rm eq}}}=\frac{4}{I^A_3}+\frac{4}{I^B_3}-\frac{48}{L_{{\rm eq}}^4}-\frac{672A_1}{L_{{\rm eq}}^8}.
\end{equation*}
Notably, the contribution from ${\cal K}_{1}$ shifts the nominal position relative to the ${\cal K}_{0}$-only limit. This shift manifests as a slight offset of the resonance stripe from the $k_1+k_2=2$ line in the network, consistent with the angular momentum features of the spin--orbit--spin resonance observed in Fig.~\ref{Fig2}.

From the Hamiltonian given by Eq. \eqref{Eq.ch4.2-d-spin--orbit--spin}, we can get the half-width of spin--orbit--spin resonance as follows:
\begin{equation*}\label{Eq29}
    {d_{\rm sos}}=\frac{\Delta (\dot \gamma_1+\dot \gamma_2)}{n_0}=\frac{8L_{{0}}^3}{I^A_3}\sqrt{\vert\frac{V_{\rm sos}}{M_{\rm sos}}\vert}.
\end{equation*}
Based on the resonant Hamiltonian $\cal K_{\rm sos}$, it is possible for us to plot the phase diagram of spin--orbit--spin resonance. With the motion integral at $I=2.60$, the phase portrait of spin--orbit--spin resonance can be found in Fig.~\ref{Fig5} (b). It is observed that there is a cat-eye structure in the phase portrait, with the resonant center at $\theta_{1}=0$ and separatrix passing through the saddle point at $\theta_1 = \pi$. The half-width of the spin--orbit--spin resonance is marked by an arrow.

\subsubsection{Spin--Spin Resonance}

The spin--spin resonance can be discussed in a similar manner to the case of spin--orbit--spin resonance. 

Initially, the Hamiltonian reads
\begin{equation}\label{Eq.ch4.2-Hamiltonian-spin--spin}
\mathcal{H}=\mathcal{H}_0+\mathcal{H}_1+\mathcal{H}_2,
\end{equation}
where $\mathcal{H}_0$ and $\mathcal{H}_1$ have the same expressions as the ones in Eq.~\eqref{Eq.ch4.2-Hamiltonian-spin--orbit--spin}, but $\mathcal{H}_2$ is different and given by
\begin{equation*}
\mathcal{H}_2=-\frac{B_6}{L^{10}}  \cos(2 \gamma_1 - 2 \gamma_2).
\end{equation*}
Through Lie-series transformation, we average out those short-period perturbation terms and obtain the spin--spin resonant Hamiltonian as follows:
\begin{equation}
\mathcal{K}_{\rm ss}=\mathcal{K}_0+\mathcal{K}_1+V_{\rm ss}\cos \theta_2,
\end{equation}
where
\begin{equation*}
\mathcal{K}_0 = -\frac{1}{2(4\Theta_1+I)^2}+\frac{2(\Theta_1-\Theta_2)^2}{I^A_3}+\frac{2(\Theta_1+\Theta_2)^2}{I^B_3},
\end{equation*}
and
\begin{equation}
\mathcal{K}_1=-\frac{A_1}{(4\Theta_1+I)^6}.
\end{equation}
The expression of $V_{\rm ss}$ is provided in Eq.~\eqref{Appendix V s-s} in Appendix \ref{A1}.

Next, we employ the canonical transformation from Eq.~\eqref{eq.ch4.2-canonical-transformation-spin--orbit--spin}. Here, the variables associated with spin--spin resonance are $(\Theta_2,\theta_2)$. Expanding the Hamiltonian at the nominal position of spin--spin resonance, the pendulum approximation can be made as follows:
\begin{equation}\label{Eq.ch4.3-d-spin--spin}
\Delta \mathcal{K}_{\rm ss} = \frac{1}{2}M(\delta\Theta_2)^2+V_{\rm ss}\cos\theta_2,\\
\end{equation}
where
\begin{equation*}
M_{\rm ss} = \frac{\partial^2 (\mathcal{K}_0+\mathcal{K}_1)}{\partial \Theta_2^2}|_{\Theta_2=\Theta_{2{\rm eq}}}=\frac{4}{I^B_3}+\frac{4}{I^A_3}.
\end{equation*}
Based on the pendulum approximation of resonant Hamiltonian, we can find the half-width of spin--spin resonance in the following manner:
\begin{equation*}
 d_{{\rm ss}} = \frac{\Delta (\dot \gamma_1-\dot \gamma_2)}{n_0}=\frac{8L_{{0}}^3}{I^A_3}\sqrt{\vert\frac{V_{\rm ss}}{M_{\rm ss}}\vert}.
\end{equation*}
Similarly, we plot the phase portrait of spin--spin resonance based on ${\mathcal K}_{\rm ss}$. In particular, with the motion integral at $I=2.57$, the phase portrait of spin--spin resonance is presented in Fig.~\ref{Fig5} (c), where the half-width of the spin--spin resonance is marked by an arrow.

As noted by \cite{hou2017note}, the center of the spin--spin resonance may shift with the change of system parameters. In our study, the equilibrium angle of the resonance is governed by the sign of the parameter $V_{\rm ss}$: when $V_{\rm ss}>0$, the equilibrium lies at $\pi$ (e.g., the top two panels of Fig.~\ref{Fig2}), whereas for $V_{\rm ss}<0$, it shifts to $0$ (e.g., the middle two panels of Fig.~\ref{Fig2} and the spin--spin phase portrait in Fig.~\ref{Fig5}). In the critical regime where $V_{\rm ss}$ changes sign, the resonance morphology under certain parameters deviates from the typical cat-eye pattern, exhibiting an anomalous structure. In this regime, \citet{lei2024primary} demonstrated that the conventional pendulum approximation fails to work. Thus, our approximation made here focuses on estimating the magnitude and trend of the resonance width rather than attempting to precisely capture the detailed structure.

\subsection{Dynamics in Region $C$}
\label{sec4.3}

As discussed in Sect.~\ref{Sect3}, region $C$ exhibits dynamical characteristics distinct from the simple superposition of two single spin--orbit synchronous resonances. Considering the fact that the traditional singly resonant analysis may fail to reveal the coupled dynamics structure of this region, we adopt the technique of Poincar\'e surfaces of section. 

Given that this region is dominated by the synchronous spin--orbit resonances of both asteroids, the nonsynchronous perturbative terms in the Hamiltonian can be averaged out, yielding the approximate Hamiltonian:
\begin{equation}\label{Eq.4.4-Hamiltonian-C}
{\mathcal H}_{C} = {\mathcal H}_0+{\mathcal H}_1,
\end{equation}
where
\begin{equation*}
\mathcal{H}_0 = -\frac{1}{2L^2} + \frac{\Gamma_1^2}{2I_3^A} + \frac{\Gamma_2^2}{2I_3^B},
\end{equation*}
and 
\begin{equation*}
\begin{aligned}
\mathcal{H}_1&=-\frac{A_1}{L^6}\left(1+\frac{3}{2}e^2\right) -\frac{A_2}{L^6}\left(1-\frac{5}{2}e^2\right)\cos(2 l - 2 \gamma_1)\\
&\quad- \frac{A_3}{L^6}\left(1-\frac{5}{2}e^2\right)  \cos(2 l - 2 \gamma_2) .
\end{aligned}
\end{equation*}
For convenience, we introduce the following canonical transformation:
\begin{equation}
\label{Eq.4.4-cononical-transformation-C}
    \begin{aligned}
\theta_1 &=2 l -2\gamma_1, \quad \Theta_1 = -\frac{\Gamma_1}{2},\\
\theta_2 &=2 l -2\gamma_2, \quad \Theta_2 = -\frac{\Gamma_2}{2},\\
\psi  &= l ,\quad\quad\quad\;\;\;\;\; I = \Gamma_1 + \Gamma_2 + L.
    \end{aligned}
\end{equation}
It is evident that the angular variable $\psi$ is a cyclic coordinate, and thus its action $I$ becomes the integral of motion. Accordingly, this Hamiltonian model determines a 2-DOF dynamical system. In particular, $\theta_1$ and $\theta_2$ are resonant angles of synchronous spin--orbit resonances of the primary and secondary asteroids. After performing transformation, the Hamiltonian $\mathcal{K}_{C}$ is written as
\begin{equation}\label{Eq.ch4.4-Kamiltonian-C}
{\mathcal K}_{C} = {\mathcal K}_0+{\mathcal K}_1,
\end{equation}
where
\begin{equation*}
\mathcal{K}_0 = -\frac{1}{2(2 \Theta_1 + 2\Theta_2 + I)^{2}} + \frac{2\Theta_{1}^{2}}{I_{3}^{A}} + \frac{2\Theta_{2}^{2}}{I_{3}^{B}},    
\end{equation*}
and
\begin{equation*}
\begin{aligned}
\mathcal{K}_1 = & -\frac{A_1}{(2 \Theta_1 + 2\Theta_2 + I)^{6}}\left(1+\frac{3}{2}e^2\right)\\
& -\frac{A_{2}}{(2 \Theta_1 + 2 \Theta_2 + I)^{6}}\left(1-\frac{5}{2}e^2\right) \cos \theta_1\\
& -  \frac{A_{3}}{(2 \Theta_1 + 2 \Theta_2 + I)^{6}}\left(1-\frac{5}{2}e^2\right)  \cos \theta_2.    
\end{aligned}
\end{equation*}

Since the Hamiltonian ${\mathcal K}_{C}$ has two degrees of freedom, it is convenient for us to study the dynamics by constructing Poincar\'e sections. Given that the synchronous spin--orbit resonance center is at $\theta_i= 0\quad (i=1,2)$, we define the Poincar\'e section as
\begin{equation}
\label{Eq.ch4.4-section}
    \begin{aligned}
\theta_2=0,\quad \dot\theta_2>0
    \end{aligned}
\end{equation}
The states $(\theta_1,\Theta_1)$ are recorded when $\theta_2$  crosses the defined section in a positive direction.

\begin{figure*}[b]
\noindent
\begin{minipage}{\linewidth}
\raggedright
\begin{minipage}[b]{0.49\linewidth}
\centering
\includegraphics[width=\linewidth]{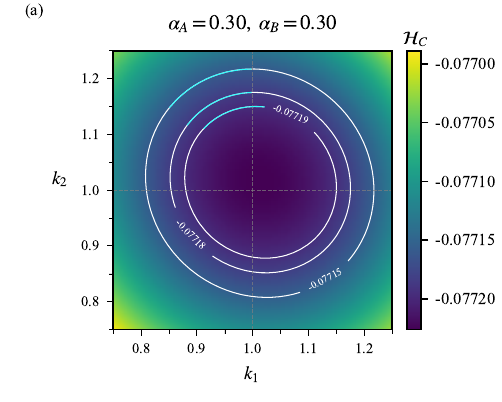}
\label{fig:heatmap}
\end{minipage}
\hfill
\begin{minipage}[b]{0.49\linewidth}
\centering
\includegraphics[width=\linewidth]{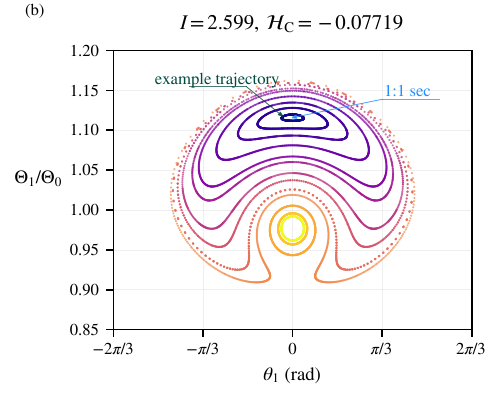}
\label{fig:poincare1}
\end{minipage}

\noindent
\begin{minipage}[b]{0.49\linewidth}
\centering
\includegraphics[width=\linewidth]{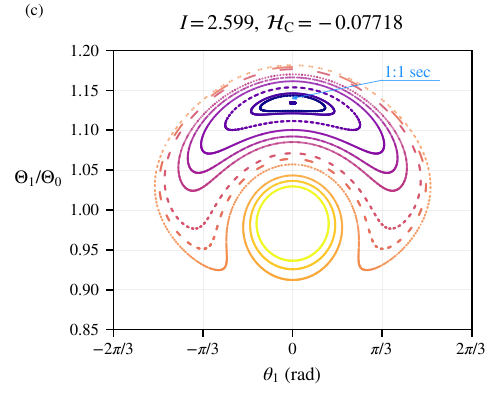}
\label{fig:poincare2}
\end{minipage}
\hfill
\begin{minipage}[b]{0.49\linewidth}
\centering
\includegraphics[width=\linewidth]{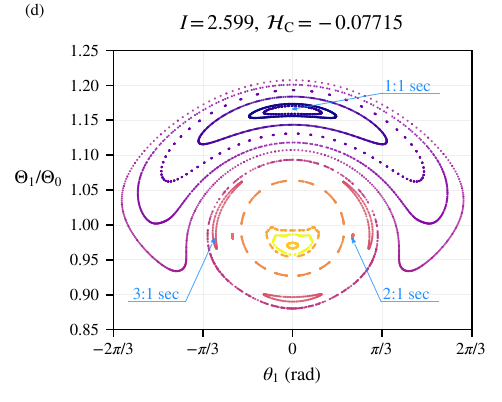}
\label{fig:poincare3}
\end{minipage}
\end{minipage}
\caption{Distribution of Hamiltonian (a) as well as Poincar\'e sections at three levels of Hamiltonian ((b)-(d)). From each of three contours shown in panel (a), we sample several initial conditions along its blue-marked segment. Then these trajectories are adopted to generate the Poincar\'e sections, as shown in panels ((b)--(d)). The remaining parameters are assumed at $\alpha_A=\alpha_B=0.3$, $G_{\rm tot}=2.596$ and $I=2.599$. The bifurcated islands corresponding to secondary resonances are marked. Evolution curves corresponding to the initial condition of the example trajectory is presented in Fig.~\ref{Fig7}.}

\label{Fig6}
\end{figure*}

We need to emphasize that, in the Poincar\'e section, all trajectories corresponding to different initial conditions should possess the same total angular momentum $G_{\rm tot}$. Meanwhile, as explained in Sect.~\ref{Sect3}, a resonant network can be constructed if the initial orbital angular momentum $L_0$ remains consistent across all points in the parameter space. Similarly, we stipulate that the total angular momentum $G_{\rm tot}$ is uniform at any grid point in the parameter space, thereby plotting the Hamiltonian distribution in the parameter space (see Fig.~\ref{Fig6} a). Then, we select three families of binary asteroids that share a common $G_{\rm tot}$. Within each family, the Hamiltonian is constant, while members differ in their initial conditions: $L_0$, $\Gamma_{10}$, and $\Gamma_{20}$.

\begin{figure}
    \centering
    \includegraphics[width=1\linewidth]{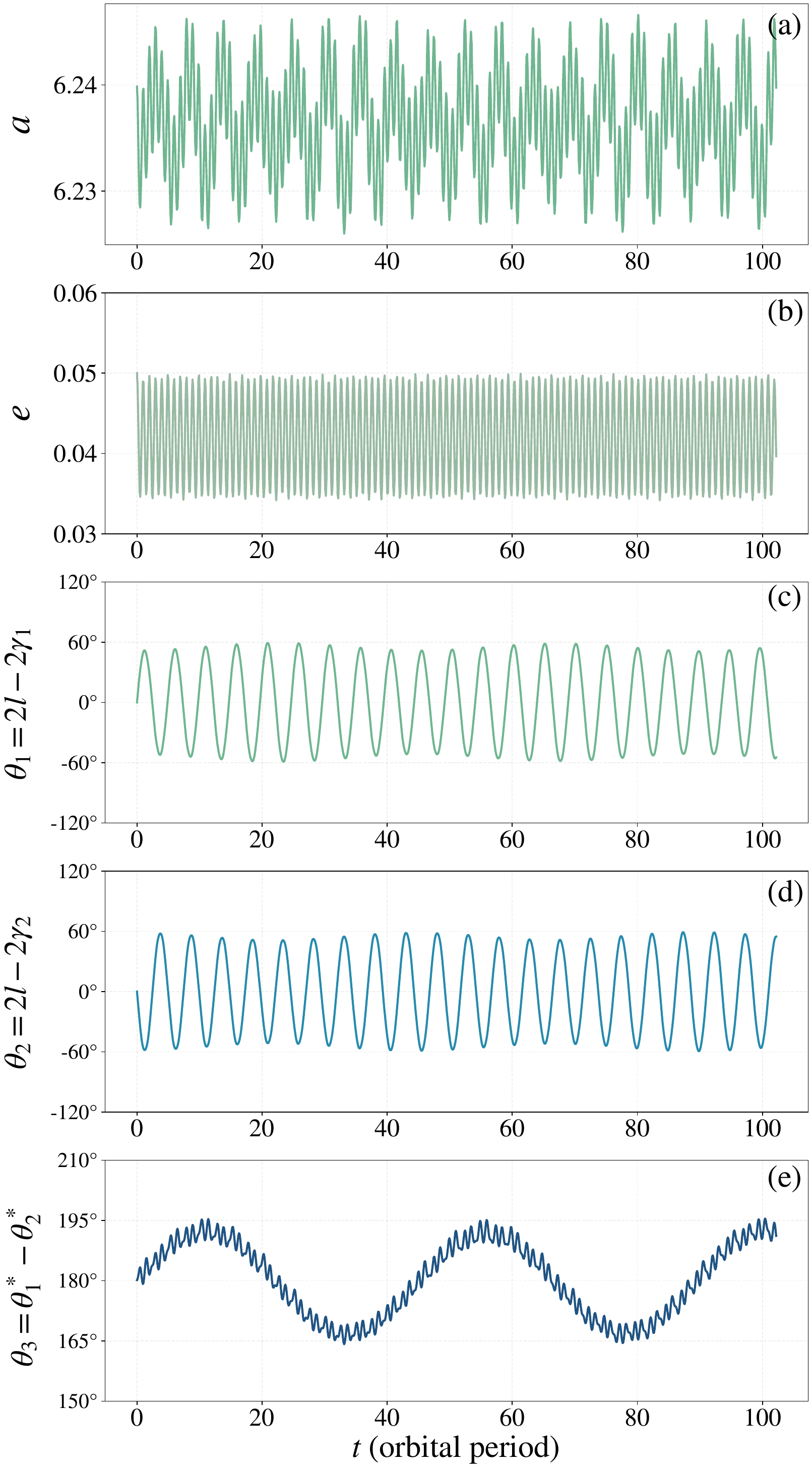}
       \caption{Evolution of semimajor axis, eccentricity and three arguments for the example trajectory corresponding to the 1:1 secondary resonance inside the doubly synchronous region (see the Poincar\'e section shown in the top right panel of Fig.~\ref{Fig6}).}
\label{Fig7}
\end{figure}

By numerically integrating their equations of motion, we plot Poincar\'e sections in Fig.~\ref{Fig6} (b)--(d) with three levels of Hamiltonian. It is observed that, across all Poincar\'e sections, an additional island of libration appears above the center of the 1:1 spin--orbit resonance. By comparing its center with its coordinates in the parameter space, we confirm that this structure corresponds to the stripe structure with a slope of $\sim$$-1$ observed in region $C$ of Fig.~\ref{Fig4}, which corresponds to the 1:1 secondary resonance.\footnote{The 1:1 secondary resonance discussed here, arising from the coupling of doubly synchronous resonances in binary asteroid systems, is distinct in its physical origin from the traditional 1:1 secondary resonance phenomena studied in, e.g., \citet{gkolias2019accurate} and \citet{callegari2024hamiltonian}. In those studies, the resonance is linked to a ratio of the free and forced frequencies at $\epsilon = \omega/n \approx 1$.} This resonance is squeezed by the underlying 1:1 spin--orbit resonance center, thereby twisting the originally cat-eye-shaped structure into a manta ray morphology. As the Hamiltonian further increases, the separation between the secondary resonance and the synchronous spin--orbit resonance center grows. Similarly, finer structures emerge in the vicinity of the resonance center, such as the 2:1 and 3:1 secondary resonances (see the bottom right panel of Fig.~\ref{Fig6}).

The Poincaré sections and the resonant network reveal the emergence of a stable 1:1 secondary resonance arising from the coupling of the doubly synchronous spin--orbit resonances. This resonance is characterized by the critical argument $\theta_3 = \theta_1^* - \theta_2^*$,
where $\theta_1^*$ and $\theta_2^*$ denote the phase-space rotation angles associated with the synchronous spin--orbit resonant arguments $\theta_1 = 2l - 2\gamma_1$ and $\theta_2 = 2l - 2\gamma_2$, respectively. These phase-space rotation angles describe the periodic evolution of the system in the $(\theta_1,\,\Theta_1)$ and $(\theta_2,\,\Theta_2)$ phase spaces. The resonance is activated when the resonant angular velocities satisfy $\dot{\theta}^{*}_{10} - \dot{\theta}^{*}_{20} \approx 0$, locking $\theta_3$ in a libration about $180^{\circ}$.

In region $C$, this resonance condition manifests as a narrow resonance valley with a slope of $\sim$$-1$. The dynamics within this valley are governed by the  
two parent spin--orbit resonances, leading to a stable secondary resonance, i.e., a resonance nested within the synchronous resonances.

Taking an example trajectory inside the secondary resonant island, we present the variations of the semimajor axis, eccentricity and critical arguments $\theta_{1,2,3}$ in Fig.~\ref{Fig7}. It is observed that there are periodic oscillations of semimajor axis and eccentricity. In addition, the three arguments $\theta_{1,2,3}$ are indeed librating, indicating that the system is confined to a secondary resonance inside the doubly synchronous region. Notably, $\theta_1$ and  $\theta_2$ evolve in an antisymmetric manner with synchronized libration amplitudes, while their motion maintains a highly synchronous and stable common period. This feature contributes to the long-term stability of the secondary resonance.

\begin{figure}
\centering
\includegraphics[width=1\linewidth]{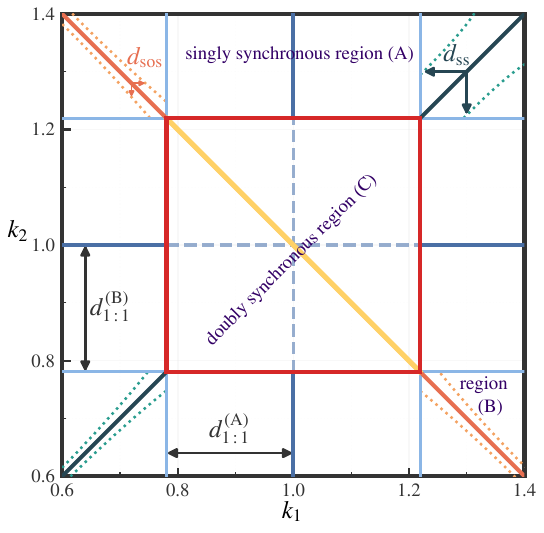}
\caption{A summary of the dynamical analysis inside the region of interest. This region is classified into three types of domains: singly synchronous spin--orbit domain $A$, nonsynchronous domain $B$, and doubly synchronous domain $C$. Inside domain $A$, half-widths of the spin--orbit synchronous resonance for the primary and secondary asteroids are denoted by $d^{A}_{\rm 1:1}$ and $d^{B}_{\rm 1:1}$. Inside the nonsynchronous region $B$, there are spin--orbit--spin and spin--spin resonances, with half-widths denoted by $d_{\rm sos}$ and $d_{\rm ss}$. Region $C$, as the doubly synchronous domain, is outlined by a red rectangle, where a stripe structure with slope of $\sim$$-1$ is caused by the 1:1 secondary resonance.}
\label{Fig8}
\end{figure}

\subsection{Summary and Discussions}

A summary of the analytical results regarding the dynamics inside the interested region is presented in Fig.~\ref{Fig8}. It presents the partition of synchronous spin--orbit resonance domains, based on the results derived in Sections~\ref{sec4.1}--\ref{sec4.3}. The centers and boundaries of the synchronous spin--orbit resonance for the primary and secondary asteroids are marked by dark blue solid lines and light blue, respectively. Their widths, including $d_{1:1}^{A}$ and $d_{1:1}^{B}$, are determined by the pendulum-approximation formula \eqref{Eq.ch4.1-d-spin--orbit}. The centers and boundaries of spin--orbit--spin and spin--spin resonances are indicated by solid and dashed lines in orange and dark green, with their corresponding half-widths $d_{\rm sos}$ and $d_{\rm ss}$ given by the analytical expressions \eqref{Eq.ch4.2-d-spin--orbit--spin} and \eqref{Eq.ch4.3-d-spin--spin}. Furthermore, the doubly synchronous domain $C$ is outlined by a red rectangle, with the center of newly identified 1:1 secondary resonances valley highlighted by a bright yellow line. The binary asteroid system on this line resides in a secondary-resonance state under the strong coupling of the doubly synchronous resonances.

The analytical results summarized in Fig.~\ref{Fig8} decipher the dynamics governing each subdomain of the interested region. This partition, initially identified in the numerical network, is thus refined analytically, yielding a unified framework that delineates the multiple resonances and their mutual boundaries. By establishing this framework, we provide a methodology grounded in the resonant network. This methodology not only allows for the a priori delineation of domains where distinct resonances govern the dynamics—with the verification of these domains provided by subsequent analysis—but also pinpoints new regions of complex dynamics, thereby guiding the discovery of novel dynamics.

It should be mentioned that, in this study, two distinct angular momentum formulations are employed, each suited to a specific stage of the investigation.

Formulation I: Fixed initial orbital angular momentum. This formulation is applied during the construction of the resonant network (Sect.~\ref{Sect3}) and the calculation of resonance widths via the pendulum approximation. It assigns a common initial orbital angular momentum $L_{0}$ to all systems, while the total angular momentum $G_{\rm tot}$ varies with the grid coordinates $(k_{1},k_{2})$.

Formulation II: Fixed total angular momentum. When plotting Poincar\'e sections (Sect.~\ref{sec4.3}), this formulation is used to obtain series of families of binary asteroids with consistent total angular momentum. Here, the coordinates of the parameter grid share the same $G_{\text{tot}}$.

Formulation I ensures that all systems in the parameter space share the same initial semi-major axis $a_0$, providing a characteristic evolutionary timescale and facilitating numerical scanning. Formulation II, on the other hand, fulfills the requirement for constructing the Poincar\'e section by assuming a fixed $G_{\rm tot}$. The two formulations are interconnected via the coordinates $(k_{1},k_{2})$ in the parameter space.

\begin{deluxetable}{lccccc}
\tablecaption{Parameters of the (90) Antiope binary asteroid system\label{tab:antiope_params}}
\tablehead{
\colhead{} & \colhead{$a$} & \colhead{$b$} & \colhead{$c$} & \colhead{$a_{\text{orb}}$} & \colhead{Size Ratio} \\
\colhead{Asteroid} & \colhead{(km)} & \colhead{(km)} & \colhead{(km)} & \colhead{(km)} & \colhead{($B/A$)}
}
\startdata
A (Primary)   & 46.5 & 43.5 & 41.8 & \multirow{2}{*}{171} & \multirow{2}{*}{0.95 $\pm$ 0.01} \\
B (Secondary) & 44.7 & 41.4 & 39.8 & & \\
\enddata
\tablenotetext{}{Notes: The data in this table come from \citet{descamps2007Antiope}. Under the assumption of equal densities for both asteroids, the size ratio $B/A = 0.95 \pm 0.01$ directly leads to the mass ratio $m_B/m_A$.}\label{table1}
\end{deluxetable}

\section{Application to (90) Antiope}
\label{Sect5}

\begin{figure*}
    \centering
    \begin{minipage}[b]{0.49\linewidth}
\centering
\includegraphics[width=\linewidth]{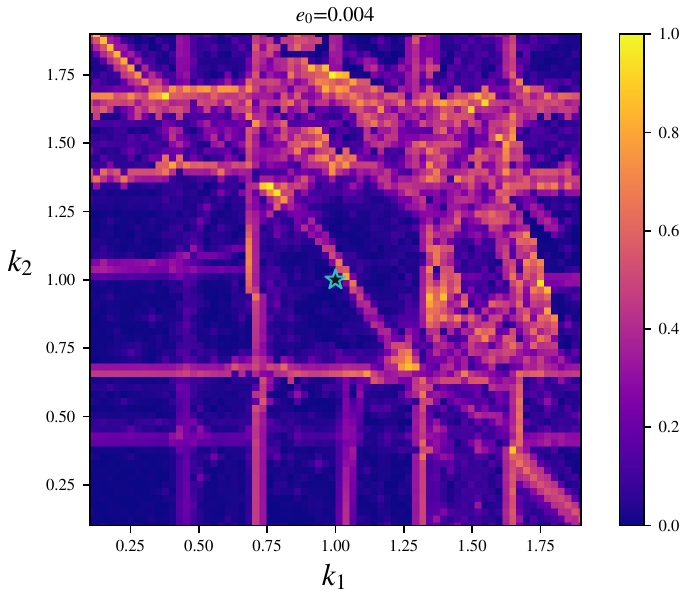}
\end{minipage}
\begin{minipage}[b]{0.49\linewidth}
\centering
\includegraphics[width=\linewidth]{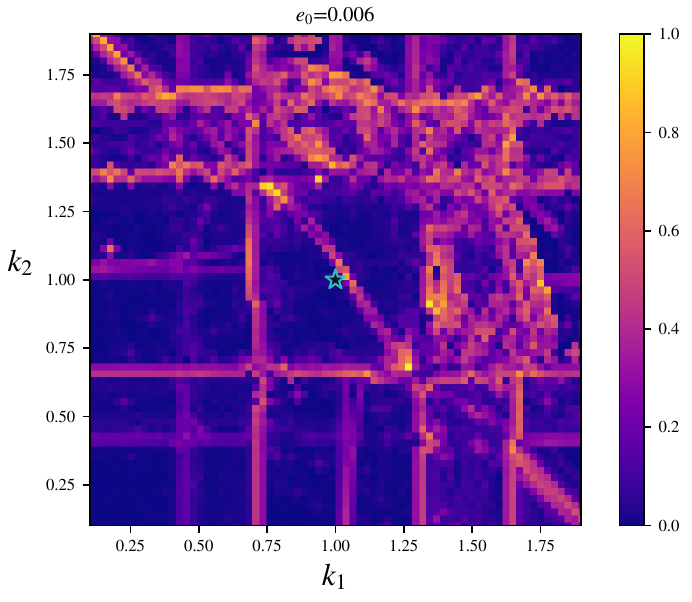}
\end{minipage}
    \caption{Resonant network of spin--orbit coupling for (90) Antiope, with initial eccentricity at $e_0=0.004$ (\textit{left panel}) and $0.006$ (\textit{right panel}). Here we adopt $\Vert \Delta a\Vert$ to characterize the resonance structure. The possible location of (90) Antiope is marked by a blue star at the point of $(k_1=1, k_2=1)$.}
    \label{Fig9}
\end{figure*}

(90) Antiope is a binary asteroid system, characterized by a relatively small orbital semi-major axis and highly symmetric primary and secondary asteroids \citep{descamps2007Antiope}. Please refer to Table~1 for the system parameters. Considering the uncertainty of orbit eccentricity, two cases are considered: $e_0=0.004$ and $e_0=0.006$. With the physical parameters given in Table~1, the resonant networks are shown in Fig.~\ref{Fig9} with $\Vert \Delta a\Vert$ as the dynamical indicator.

In general, we can see that (90) Antiope holds similar global structures to the ones shown in Fig.~\ref{Fig2}. Thus, the structures shown in Fig.~\ref{Fig9} can be understood with the help of the analysis made in Sect.~\ref{Sect4}. 

In the central region, there is a doubly synchronous zone, which is dominated by synchronous spin--orbit resonances of both asteroids. Inside the regime, we observe a prominent stripe corresponding to the secondary spin--orbit resonance discussed in Sect.~\ref{Sect4}. Due to a slight asymmetry in shape parameters between two asteroids ($\alpha_A=0.45,\;\alpha_B=0.48$), the secondary resonance manifests as a stripe aligned along the diagonal of the rectangular doubly synchronous region, with a slope slightly offset from $-1$, in contrast to the spin--orbit--spin resonance, which retains a standard slope of $-1$. In addition, the structures appeared outside the doubly synchronous domain are governed by singly spin--orbit resonances and spin--orbit--spin resonances. While the spin--orbit--spin structures adjacent to the doubly synchronous region are obscured by chaos, clear signatures persist in the upperleft and lowerright corners of the network. However, it should be noted that the spin--spin resonant structure is not evident in Fig.~\ref{Fig9} because the adopted dynamical indicator $\Vert \Delta a\Vert$ is not sensitive for the spin--spin resonance.

\section{conclusion}
\label{Sect6}

In this work, we investigate the global dynamical structures of spin--orbit coupling in the $(k_1,k_2)$ parameter space for ellipsoid–ellipsoid binary asteroid systems. In particular, we conduct a systematic study of the dynamics in the vicinity of the synchronous spin--orbit resonances of both asteroids. The analytical results provide insight into the underlying dynamical mechanisms governing resonant networks.

By adopting a second-derivative-based index as a dynamical indicator, we construct a series of resonant networks for binary asteroid systems with different combinations of shape parameters $(\alpha_A,\alpha_B)$. This approach reveals the global dynamical structure, showing that the dominant features are governed by the spin--orbit resonances of the primary and secondary asteroids. In particular, the 2:1, 1:1, and 2:3 spin--orbit resonances of both bodies form an interconnected network of resonances in the considered parameter space. Notably, a distinctive stripe-like structure with an approximate slope of $-1$ is observed, extending from the upper left corner to the lower right corner of the parameter space. 

To elucidate the dynamics of the resonant networks, we focus on the parameter space in the vicinity of the synchronous spin--orbit resonances of both asteroids. Based on the widths of the individual synchronous resonances, this region is divided into three types of domains: the singly synchronous domain ($A$), the nonsynchronous domain ($B$), and the doubly synchronous domain ($C$). The dynamics in domain ($A$) are governed by the spin--orbit synchronous resonance of either asteroid. In domain ($B$), the dynamics are dominated by either spin--orbit--spin resonances or spin--spin resonances. The dynamics in domain ($C$) are controlled by the synchronous resonances of both asteroids.

Specifically, expressions of resonant Hamiltonian are derived for the single spin--orbit synchronous resonance, as well as for the spin--spin and spin--orbit--spin resonances, using perturbation theory. Explicit expressions for the half-width of each resonance are obtained under the pendulum approximation. Within the doubly synchronous domain, the resulting Hamiltonian defines a 2-DOF dynamical model, which is analyzed using Poincar\'e sections. The results show that the stripe structure with a slope of $-1$ outside the synchronous region is induced by the spin--orbit--spin resonance, whereas the stripe structure inside the synchronous region is generated by a secondary 1:1 resonance stemming from the coupling of doubly synchronous spin--orbit resonances. 
In addition, the faint stripe structure with a slope of $+1$ is attributed to the spin--spin resonance.

An application to the binary asteroid system (90) Antiope demonstrates that the parameter space $(k_1,k_2)$ exhibits a rich and intricate dynamical structure. By applying the analytical framework developed in this work, the complex global patterns arising within the resonant network can be systematically interpreted. In particular, the identified resonant mechanisms provide a clear explanation for the origin, interaction, and organization of the dominant structures observed in the parameter space, thereby offering a coherent dynamical understanding of the system.

In summary, this work establishes a unified framework that bridges global numerical mapping and local analytical verification to systematically partition the parameter space of binary asteroid systems into distinct dynamical domains, including 1:1 spin--orbit, spin--orbit--spin, spin--spin, and doubly synchronous spin--orbit. 
Although the present study concentrates on the synchronous region, the analytical methodology and dynamical insights developed here are readily applicable to a much broader parameter space. In particular, they can be naturally extended to regions associated with arbitrary pairs of spin--orbit resonances within the full resonant network, providing a general tool for understanding the complex spin--orbit dynamics of binary asteroid systems.


\appendix

\section{Explicit expressions of $W_1$, $V_{\rm sos}$ and $V_{\rm ss}$}
\label{A1}

The generating function $W_1$ is given by
\begin{equation}\label{W1}
\begin{aligned}
W_1 & = -\frac{3A_1}{L^3}e\sin(l)+X_1A_2\sin(l-2\gamma_1)+X_2A_2\sin(2l-2\gamma_1)+X_3A_2\sin(3l-2\gamma_1)+Y_1A_3\sin(l-2\gamma_2)\\
&+Y_2A_3\sin(2l-2\gamma_2)+Y_3A_3\sin(3l-2\gamma_2)
\end{aligned}
\end{equation}
where
\begin{equation*}
\begin{aligned}
X_1 &= \frac{e}{2L^6} \left( \frac{1}{L^3} - 2\frac{\Gamma_1}{I^A_3} \right)^{-1},\quad
X_2 = -\frac{1}{2L^6} \left( \frac{1}{L^3} - \frac{\Gamma_1}{I^A_3} \right)^{-1},\quad
X_3 = -\frac{7e}{2L^6} \left( \frac{3}{L^3} - 2\frac{\Gamma_1}{I^A_3} \right)^{-1},\\
Y_1 &= \frac{e}{2L^6} \left( \frac{1}{L^3} - 2\frac{\Gamma_2}{I^B_3} \right)^{-1},\quad
Y_2 = -\frac{1}{2L^6} \left( \frac{1}{L^3} - \frac{\Gamma_2}{I^B_3} \right)^{-1},\quad
Y_3 = -\frac{7e}{2L^6} \left( \frac{3}{L^3} - 2\frac{\Gamma_2}{I^B_3} \right)^{-1}.
\end{aligned}
\end{equation*}

The coefficient $V_{\rm sos}$ in the spin--orbit--spin resonant Hamiltonian is
\begin{equation}\label{Appendix V s-o-s}
\begin{aligned}
V_{\rm sos} &= -\frac{B_7}{L^{10}} + \frac{3A_2 A_3}{4L^{10}} \left[ 
\frac{(I_3^A)^2}{(I_3^A - L^3\Gamma_1)^2} + \frac{(I_3^B)^2}{(I_3^B - L^3\Gamma_2)^2} \right] + \frac{7A_2 A_3}{16L^{10}} \Bigg[ 
\frac{I_3^A\left[(-2 + 5e^2)I_3^A + 4(1 - 7e^2)L^3\Gamma_1\right]}{(I_3^A - 2L^3\Gamma_1)^2} \\
&+ \frac{I_3^A\left[(6 - 51e^2)I_3^A + 4(-1 + 7e^2)L^3\Gamma_1\right]}{(3I_3^A - 2L^3\Gamma_1)^2}+ \frac{I_3^B\left[(-2 + 5e^2)I_3^B + 4(1 - 7e^2)L^3\Gamma_2\right]}{(I_3^B - 2L^3\Gamma_2)^2} \\
&+ \frac{I_3^B\left[(6 - 51e^2)I_3^B + 4(-1 + 7e^2)L^3\Gamma_2\right]}{(3I_3^B - 2L^3\Gamma_2)^2} \Bigg].
\end{aligned}
\end{equation}

The coefficient $V_{\rm ss}$ in the spin--spin resonant Hamiltonian is given by
\begin{equation}\label{Appendix V s-s}
\begin{aligned}
V_{\rm ss}=& -\frac{B_6}{L^{10}} +\frac{1}{8L^{10}} A_2 A_3 \Bigg[ 
\frac{12I_3^A(3I_3^A - 4L^3\Gamma_1)}{(I_3^A - L^3\Gamma_1)^2}+ \frac{12I_3^B(3I_3^B - 4L^3\Gamma_2)}{(I_3^B - L^3\Gamma_2)^2} \\
&+ \frac{I_3^A\left[(-2 + 11e^2 + 4\sqrt{1-e^2})I_3^A - 4(-1 + 7e^2 + 2\sqrt{1-e^2})L^3\Gamma_1\right]}{(I_3^A - 2L^3\Gamma_1)^2} \\
&+ \frac{49I_3^A\left[3(-6 + 33e^2 + 4\sqrt{1-e^2})I_3^A - 4(-3 + 21e^2 + 2\sqrt{1-e^2})L^3\Gamma_1\right]}{(3I_3^A - 2L^3\Gamma_1)^2} \\
&+ \frac{I_3^B\left[(-2 + 11e^2 + 4\sqrt{1-e^2})I_3^B - 4(-1 + 7e^2 + 2\sqrt{1-e^2})L^3\Gamma_2\right]}{(I_3^B - 2L^3\Gamma_2)^2} \\
&+ \frac{49I_3^B\left[3(-6 + 33e^2 + 4\sqrt{1-e^2})I_3^B - 4(-3 + 21e^2 + 2\sqrt{1-e^2})L^3\Gamma_2\right]}{(3I_3^B - 2L^3\Gamma_2)^2} \Bigg].
\end{aligned}
\end{equation}

\section{Resonant networks in different configurations}
\label{A2}

\begin{figure}
    \centering
    \begin{minipage}[b]{0.31\linewidth}
\centering
\includegraphics[width=\linewidth]{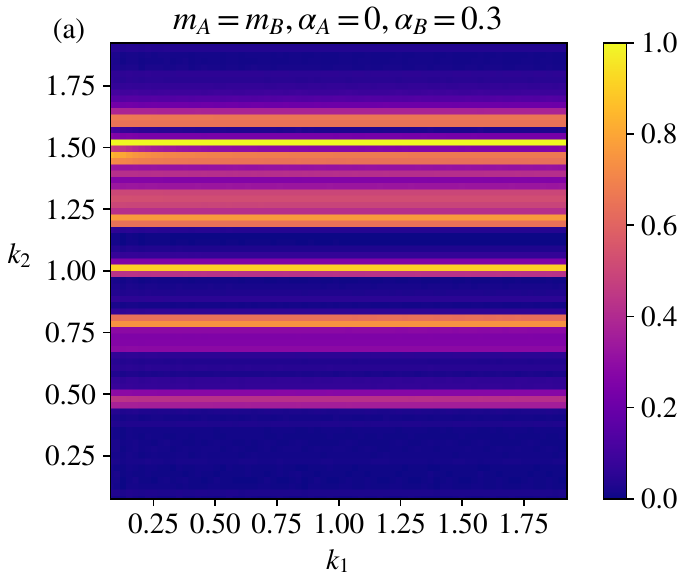}
\end{minipage}
\centering
    \begin{minipage}[b]{0.31\linewidth}
\centering
\includegraphics[width=\linewidth]{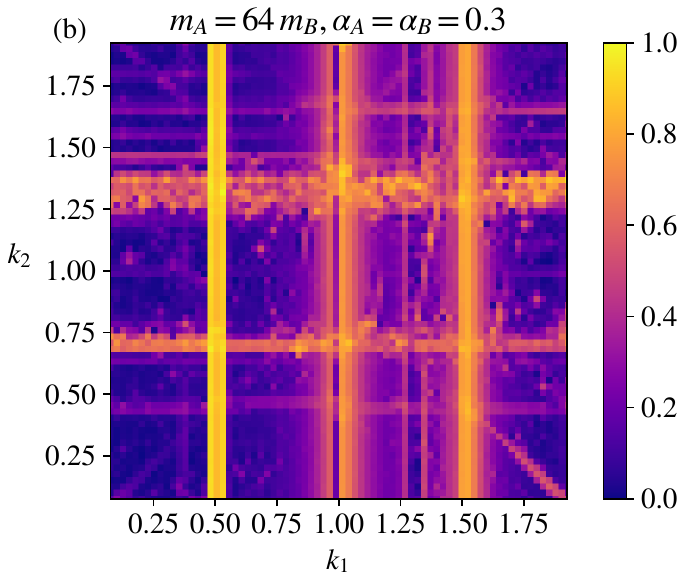}
\end{minipage}
\begin{minipage}[b]{0.31\linewidth}
\centering
\includegraphics[width=\linewidth]{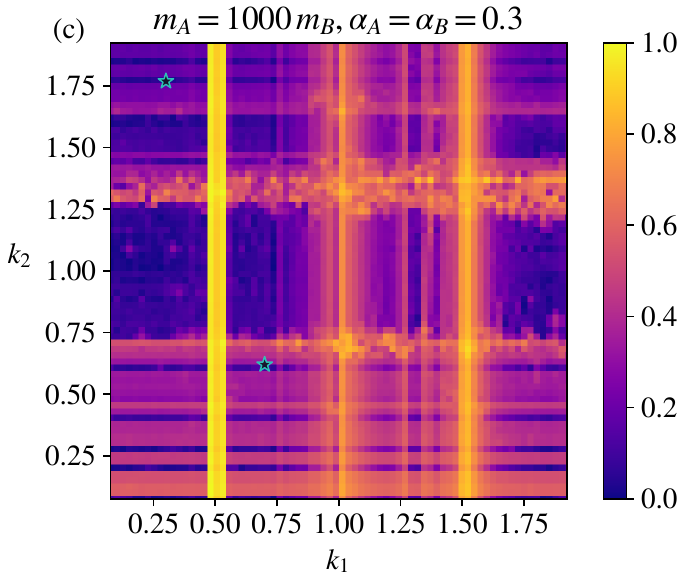}
\end{minipage}
    \caption{Resonant networks for (a) a spherical primary ($\alpha_A=0, \alpha_B=0.3, m_A = m_B$), (b) $m_A = 64\,m_B$ ($\alpha_A = \alpha_B = 0.3$), and (c) $m_A = 1000\,m_B$ ($\alpha_A = \alpha_B = 0.3$), all with initial eccentricity $e_0 = 0.05$. In panel (c), two special points on the spin--orbit--spin resonance stripe ($k_1$ = 0.300, $k_2$ = 1.768) and on the spin--spin resonance stripe ($k_1$ = 0.700, $k_2$= 0.620) are marked by blue stars, with their associated trajectories shown in Fig.~\ref{figA4-5}.}
    \label{figA1-3}
\end{figure}

To broaden the scope of our analysis, we present the resonant network for the case of a spherical primary ($\alpha_A=0$) and ellipsoidal secondary ($\alpha_B=0.3$), as shown in panel (a) of Fig.~\ref{figA1-3}. In this configuration, all the spin--spin and spin--orbit--spin resonances, as well as the 1:1 secondary resonance, disappear because the spin--orbit coupling of the primary asteroid vanishes. Consequently, the phase space is dominated by the spin--orbit resonance of the secondary asteroid (horizontal stripes are evident).

\begin{figure}
    \centering
    \begin{minipage}[b]{0.4\linewidth}
\centering
\includegraphics[width=\linewidth]{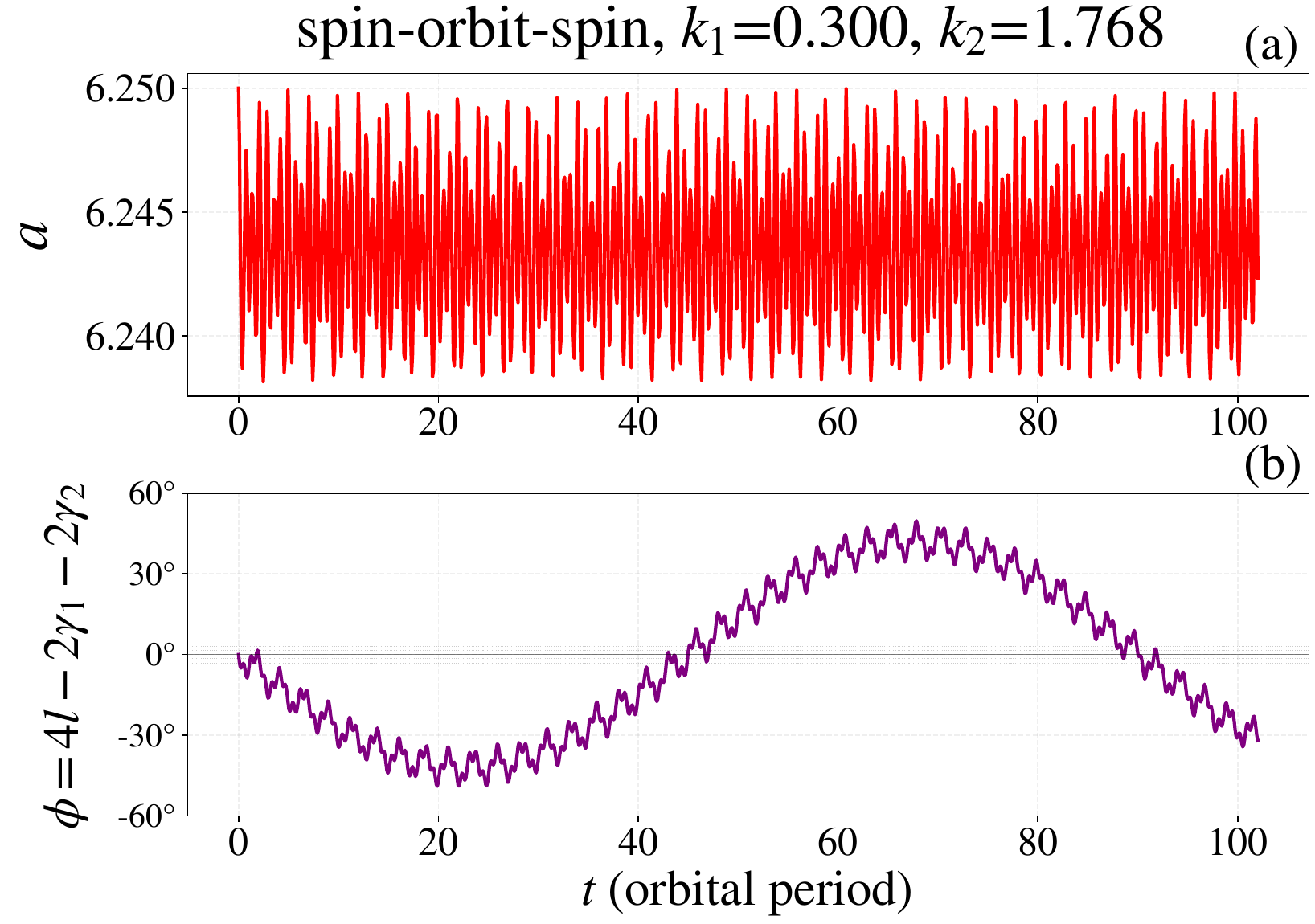}
\end{minipage}
\begin{minipage}[b]{0.4\linewidth}
\centering
\includegraphics[width=\linewidth]{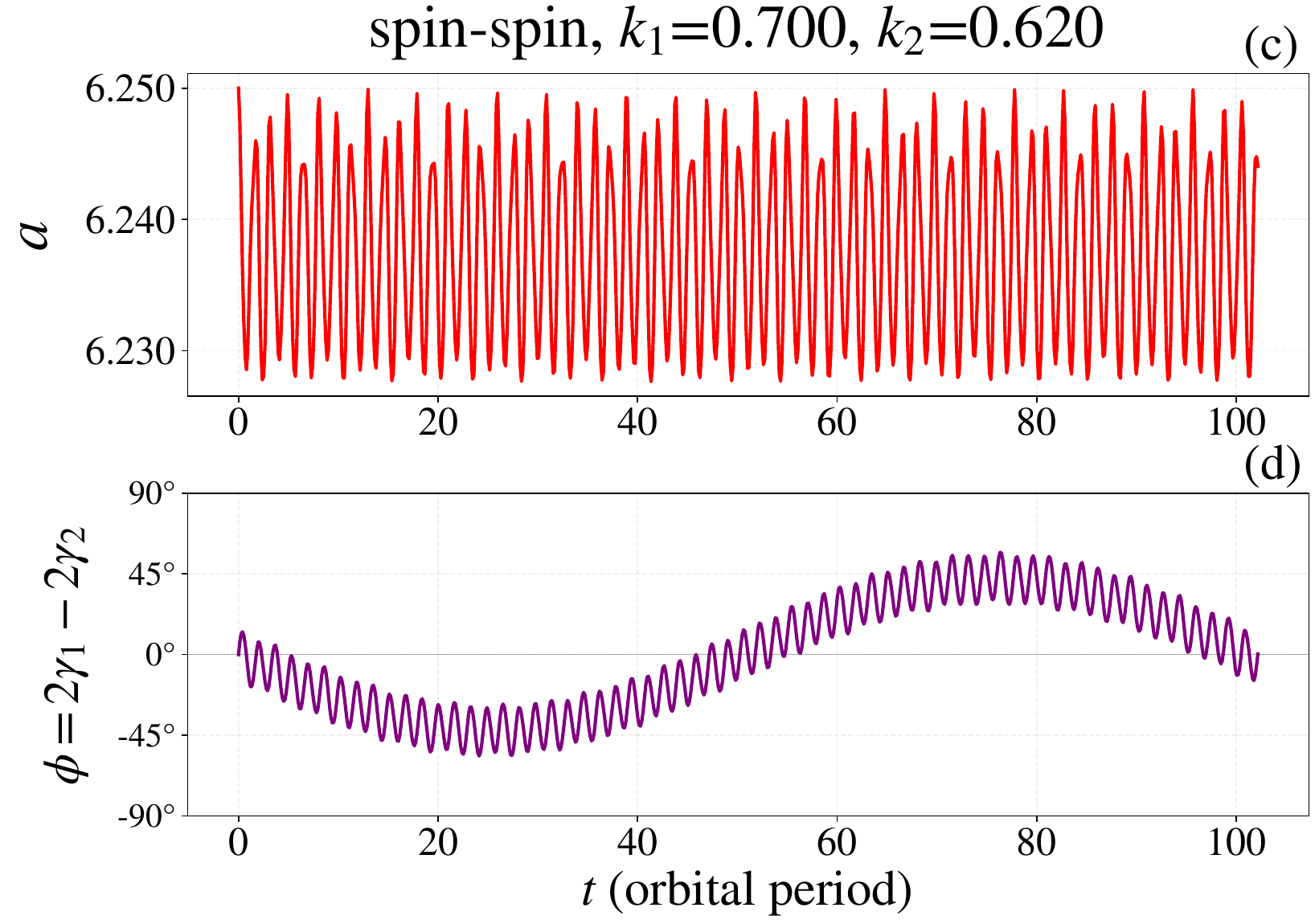}
\end{minipage}
    \caption{Time evolution of semimajor axis and critical arguments for the resonant trajectories with $m_A=1000\,m_B,\,\alpha_A=\alpha_B= 0.3$.  The left panels  are for the spin--orbit--spin resonance under the condition of $k_1=\,0.300,\;k_2 = 1.768$, and the right panels are for the spin--spin resonance under the condition of $k_1=0.700,\;k_2 = 0.620$.}
    \label{figA4-5}
\end{figure}

 In order to investigate the influence of mass asymmetry of the binary asteroid system, we perform a simulation for the case of $m_A \gg m_B$, as shown in panels (b) and (c) of Fig.~\ref{figA1-3}. When $m_A = 64\,m_B$, bright stripes corresponding to spin--orbit--spin and spin--spin resonances can be observed, whereas when $m_A = 1000\, m_B$, the spin--orbit--spin and spin--spin resonance structures are no longer visible. Furthermore, compared to mass-symmetric  asteroids' network, the width of the primary's 1:1 synchronous resonance decreases significantly, whereas that of the secondary increases. This behavior can be explained by the theoretical analysis ( see Eq.~\eqref{Eq.K11} and Eq.~\eqref{Eq.ch4.1-d-spin--orbit}).

While spin--spin and spin--orbit--spin resonant structures are not visible in the  network when $m_A=1000\,m_B$, we emphasize that they do not really vanish. Through detailed study, we have successfully identified specific initial conditions that can excite spin-spin and spin-orbit-spin resonances in the case of $m_A \gg m_B$. See Fig.~\ref{figA4-5}  for the corresponding evolution curves. We attribute the absence of these resonances in the global network to their extremely weak dynamical indicator compared to the dominant spin--orbit resonances. Furthermore, when the initial conditions of the binary asteroid system satisfy the excitation criteria for both spin--orbit resonance and spin-orbit-spin (or spin--spin) resonance, the weaker resonance may be disrupted under the dominant influence of the stronger spin-orbit resonance, preventing it from maintaining long-term stability.

\bibliography{mybib}{}
\bibliographystyle{aasjournal}



\end{document}